\documentclass[aps,prd,floatfix,superscriptaddress,preprintnumbers,nofootinbib,twocolumn,longbibliography]{revtex4-2}
\pdfoutput=1

\usepackage{amsmath, amssymb, amsfonts, graphicx, mathrsfs, verbatim, float, slashed, bbm, color, xcolor, xspace, enumerate, url, bbding, multirow, outlines, upgreek}
\usepackage[english]{babel}
\usepackage{comment}
\usepackage[version=4]{mhchem} 
\usepackage{tikz}
\usepackage{hyperref}

\definecolor{lightsabergreen}{rgb}{.14,.64,.14}

\hypersetup{
     colorlinks   = true,
     citecolor    = lightsabergreen,
     urlcolor     = lightsabergreen,
     linkcolor    = lightsabergreen
}

\definecolor{lime}{HTML}{A6CE39}
\DeclareRobustCommand{\orcidicon}{%
	\begin{tikzpicture}
	\draw[lime, fill=lime] (0,0) 
	circle [radius=0.16] 
	node[white] {{\fontfamily{qag}\selectfont \tiny ID}};
	\draw[white, fill=white] (-0.0625,0.095) 
	circle [radius=0.007];
	\end{tikzpicture}
	\hspace{-3mm}
}

\foreach \x in {A, ..., Z}{%
	\expandafter\xdef\csname orcid\x\endcsname{\noexpand\href{https://orcid.org/\csname orcidauthor\x\endcsname}{\noexpand\orcidicon}}
}

\begin{document}

\author{Javier F. Acevedo}
\thanks{\href{mailto:jfacev@uvic.ca}{jfacev@uvic.ca}}
\affiliation{Particle Theory Group, SLAC National Accelerator Laboratory, Stanford, CA 94035, USA}
\affiliation{Department of Physics and Astronomy, University of Victoria, Victoria, BC V8P 5C2, Canada}
\affiliation{TRIUMF, Vancouver, BC V6T 2A3, Canada}

\title{Milky Way White Dwarfs as Inelastic Dark Matter Detectors}
\begin{abstract}
We show that white dwarfs in the Galactic Center are a sensitive probe of inelastic dark matter (IDM) annihilation. IDM particles traversing a white dwarf may become gravitationally captured without fully settling in its interior if their energy drops below the kinematic threshold for inelastic scattering. In this regime, a fraction of the incoming particles remains bound on long-lived orbits that extend beyond the white dwarf's volume, enabling annihilation to partially proceed outside of the object. Some of the annihilation products subsequently escape, rather than being absorbed by the stellar material, producing a potentially detectable signal. We estimate the collective gamma-ray flux from the external annihilation sourced by these systems, as a function of the IDM parameters, and compare it to H.E.S.S. observations within the central degree of the galaxy. From this comparison, we derive tentative limits on the IDM scattering cross-section at the TeV mass scale, under largely conservative assumptions for the IDM density in this region. As a concrete application, we map these results into specific parameter constraints for dark photon-mediated and higgsino models with mass splittings at the MeV scale.
\end{abstract}

\maketitle


\section{Introduction}
Over the past decades, white dwarfs have received increasing attention as large-volume dark matter detectors. Their extreme densities and relatively large radii enable efficient capture of halo dark matter even for tiny scattering cross-sections, while their deep gravitational potentials suppress evaporation of captured particles across several orders of magnitude in dark matter mass. These combined properties allow these objects to probe a broad range of dark matter particle properties. Proposed signatures of dark matter in these objects include kinetic and annihilation heating of the host white dwarf \cite{Mochkovitch:1985vi,Moskalenko:2006mk,McCullough:2010ai,Hooper:2010es,Bell:2021fye,Garani:2023esk,Leane:2024bvh,HoefkenZink:2024hor,Acevedo:2024zkg,Zhang:2024qof, HoefkenZink:2026nrh, Chu:2026mwo}, detectable fluxes of Standard Model (SM) particles from dark matter annihilation into decaying intermediate states \cite{Dasgupta:2019juq,Panotopoulos:2020kuo,DeRocco:2022rze,Ramirez-Quezada:2022uou,Acevedo:2023xnu,Leane:2024bvh,Bhattacharjee:2025iip}, and the ignition of Type-Ia supernovae in white dwarfs near the Chandrasekhar limit \cite{Graham:2015apa,Bramante:2015cua,Krall:2017xij,Graham:2018efk,Acevedo:2019gre,Janish:2019nkk,Fedderke:2019jur,Curtin:2020tkm,Acevedo:2020avd,Acevedo:2021kly,Raj:2023azx,Acevedo:2023cab,Jiang:2025xln}.

Recently, it was shown that white dwarfs in the inner Milky Way can be a sensitive probe of dark matter annihilation, for models in which annihilation produces long-lived or boosted particles that escape these objects and decay outside into visible states \cite{Acevedo:2023xnu}. Near the Galactic Center, where the underlying dark matter density is high, these objects capture and thermalize large amounts of dark matter, which can then annihilate at an enhanced rate relative to the halo. Since capture and annihilation are expected to rapidly equilibrate over the white dwarf's lifetime, comparing the resulting collective flux against, $e.g.$, gamma-ray observations of this region directly constrains the dark matter's scattering cross-section. 

While highly sensitive, this search crucially requires the existence of intermediate states that ensure the released energy escapes the host object, as otherwise any annihilation product (except, potentially, low-energy neutrinos) would be absorbed as unobservable heat by the stellar material. This assumption can be lifted, however, if the dark matter does not completely settle in the object's interior prior to annihilating. Such is the case, for instance, in inelastic dark matter (IDM) scenarios \cite{Tucker-Smith:2001myb,Chang:2008gd,Finkbeiner:2007kk,Pospelov:2007xh,Batell:2009vb,Ghorbani:2014gka,Zhang:2016dck,Alvarez:2019nwt,Hooper:2025fda}, where tree-level scattering involves two states of different masses $\chi_1$ and $\chi_2$. This introduces an additional kinematic threshold that can effectively prevent the IDM from fully settling inside the white dwarf. 

This partial thermalization of IDM was first explored in Ref.~\cite{Acevedo:2024ttq}, which considered Galactic Center neutron stars, and focused on annihilation proceeding outside the stellar volume. In this work, we extend the analysis to white dwarfs within the same region. Owing to their weaker gravitational field compared to neutron stars, these objects probe a more limited range of mass splitting between the light and heavy state, about 1$-$12 MeV when the IDM is heavier than the stellar constituents. However, they also offer a few decisive advantages: their larger radii allow for higher IDM capture rates and consequently stronger individual annihilation rates, while their greater abundance further boosts their collective contribution. 

\begin{figure}[t!]
    \centering
    \includegraphics[width=1.08\linewidth]{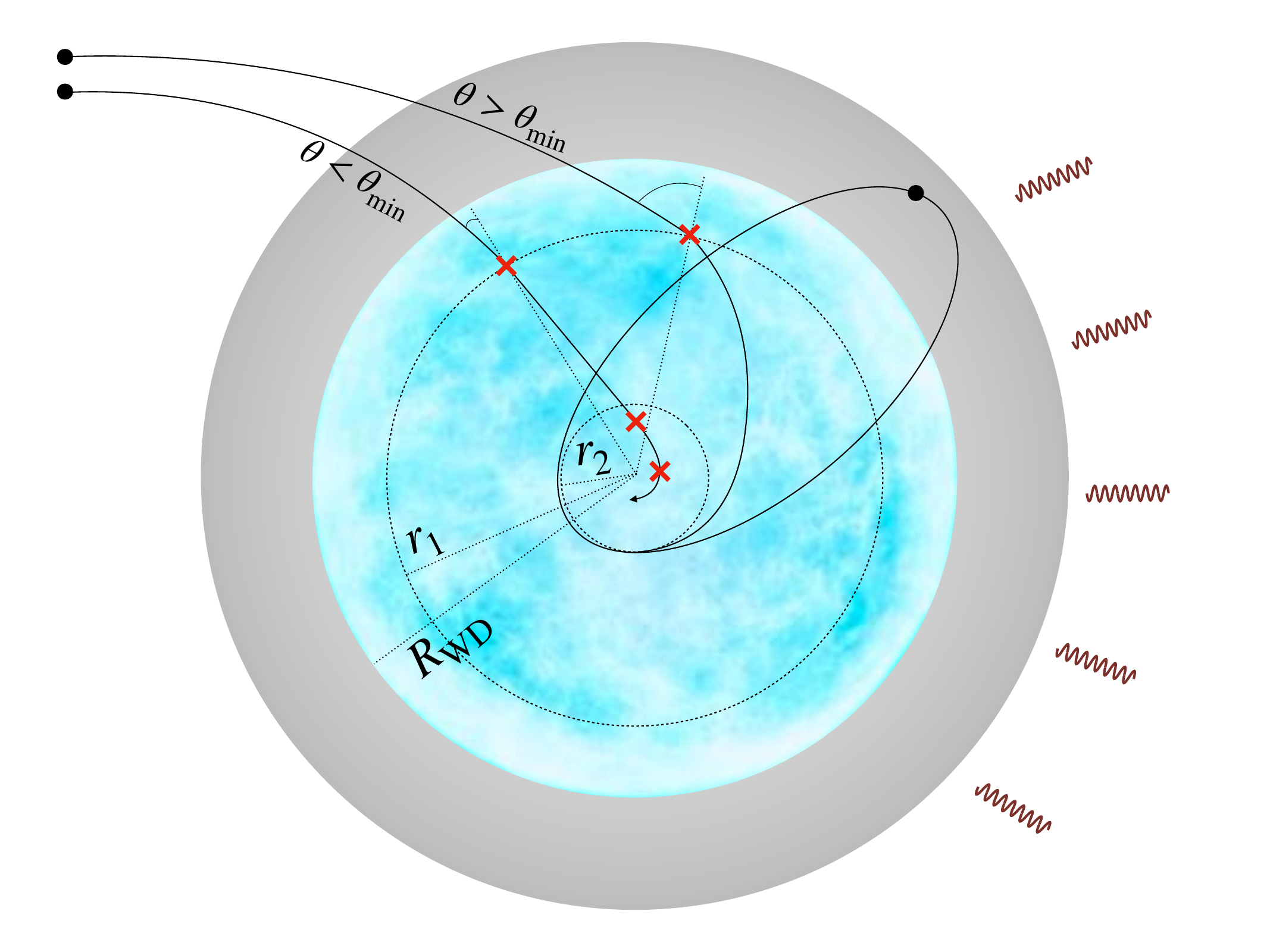}
    \caption{Schematic of heavy IDM capture and partial thermalization in a white dwarf. Endothermic inelastic scattering is initially allowed within a shell of radius $r_1$, where the local velocity exceeds the threshold set by the mass splitting. Two incoming trajectories are shown, entering the shell at angle $\theta$ relative to the radial direction. In both cases shown, a single scatter suffices to drop the particle's velocity below threshold. For $\theta < \theta_{\rm min}$, the particle penetrates deep into the shell and undergoes repeated scatters, sinking toward the center and eventually thermalizing. For $\theta > \theta_{\rm min}$, the particle does not reach the threshold velocity again after the initial energy loss, settling onto a long-lived orbit with pericenter $r_2$ and apocenter beyond $R_{\rm WD}$, and therefore permitting annihilation to proceed outside.}
    \label{fig:schem}
\end{figure}

Figure~\ref{fig:schem} shows a depiction of the external annihilation for heavy IDM: depending on the mass splitting, inelastic scattering is kinematically allowed only within a restricted region of the white dwarf's interior, where the local escape velocity is large enough to exceed the threshold required for the transition. IDM particles reaching this shell scatter almost immediately if the inelastic cross-section is sufficiently large, rapidly losing energy until they drop below the threshold velocity for inelastic scattering. From this point on, they free-fall through the white dwarf's interior. A fraction of these particles retains sufficient angular momentum that their velocity remains below threshold, even at their point of closest approach to the center where their velocity is maximized. Such particles then remain pinned onto long-lived orbits that enable external annihilation. We construct a semi-analytic framework to calculate the configuration and occupation number of these extended orbits, the associated IDM density surrounding white dwarfs, and the resulting external annihilation rate. 

We focus here on the production of gamma-rays. We compare the collective flux from white dwarfs, for various sample annihilation channels, against existing observations of the Galactic Center performed by the High Energy Stereoscopic System (H.E.S.S.), which we use to derive prospective limits on the IDM's scattering cross-section. Our semi-analytic approach is restricted to the regime where the IDM is much heavier than the stellar constituents. Combined with H.E.S.S.'s accessible energy range, we focus on an IDM mass range roughly at the TeV scale. Various well-motivated IDM models populate this mass range, and our results show that white dwarfs in this region can probe novel parameter space within it, even under conservative assumptions about the IDM profile toward the inner galaxy. Interestingly, this same parameter space is largely inaccessible to direct detection experiments: in models where no relic heavy state $\chi_2$ survives, the characteristic IDM velocity in the halo limits their sensitivity to mass splittings $\lesssim 400 \ \rm{keV}$ for endothermic transitions $\chi_1 \to \chi_2$ \cite{Bramante:2016rdh}. The absence of relic $\chi_2$ states is, in fact, generically the case for the example models with MeV mass splittings we consider. Consequently, white dwarfs can offer a powerful, complementary probe of IDM scenarios in a regime that is experimentally challenging to access.

The remainder of this paper is as follows: in Section~\ref{sec:gc_mod}, we first detail our modeling of the white dwarf population in the central parsec of the Milky Way, along with the benchmark IDM profiles we consider. In Section~\ref{sec:cap_review}, we review the inelastic threshold and how it determines the overall flux of IDM that can be captured by white dwarfs. In Section~\ref{sec:partial_therm} we review the conditions under which captured IDM particles cease to thermalize with these objects, as well as their corresponding orbital configuration. The equilibrium state between capture, annihilation, and orbital decay is analyzed in Section~\ref{sec:cap-orb-ann-eq}. In Section~\ref{sec:collective_signal}, we compute the collective, external annihilation rate by white dwarfs in this region, and compare it against H.E.S.S. observations for some example annihilation channels. We discuss concrete models for which this analysis would be applicable in Section~\ref{sec:IDM_models}. Finally, we conclude in Section~\ref{sec:conclusions}. Throughout this work, we utilize natural units where $\hbar = c = k_b = 1$ and $G = 1/M_{\rm Planck}^2$.

\section{Galactic Center Modeling}
\label{sec:gc_mod}
To compute the collective annihilation signal, we require the abundance of white dwarfs, their spatial distribution, and the underlying IDM density and velocity dispersion. As our region of interest, following Ref.~\cite{Acevedo:2023xnu}, we will consider the Milky Way's nuclear star cluster. This corresponds to the central few parsecs around the supermassive black hole Sagittarius A$^*$ (Sgr.~A$^*$), where both the stellar and dark matter densities are robustly known to be high. Although other systems, such as globular clusters, could also serve as promising targets, we do not consider them here because the evidence for significant dark matter content in these systems currently remains largely inconclusive.

\subsection{White Dwarf Properties and Distribution}
Theoretical modeling \cite{1977ApJ...216..883B,Quinlan:1994ed,1980ApJ...242.1232Y,Hopman_2006,Alexander:2008tq}, as well as N-body simulations \cite{Amaro-Seoane:2010dzj,2018A&A...609A..27S,Panamarev:2018bwq,2024ApJ...961..232Z}, strongly indicate the number density of white dwarfs in the Milky Way's nuclear star cluster follows a power-law profile. Furthermore, power-law distributions for stellar objects in this region are also favored by recent spectroscopic surveys \cite{2019ApJ...872L..15H,2018A&A...609A..26G,2018A&A...609A..28B} and observations of x-ray binaries \cite{2018Natur.556...70H}. Motivated by these results, for the white dwarf number density we adopt a simple power-law of the form 
\begin{equation}
    n_{\rm WD}(R) = n_{\rm WD}(R_0) \left(\frac{R}{R_0}\right)^{-\alpha} \, \, (R \leq R_0) ~,
    \label{eq:wd_dist}
\end{equation}
where $R$ is the galactocentric distance, $\alpha$ is the power-law index, and $n_{\rm WD}(R_0)$ determines the normalization, which we fix below. The index ranges between $1.0 - 1.4$ depending on the assumed star formation history, initial mass function, and initial-to-final mass relation for compact stars. For our calculations, we adopt $\alpha = 1.4$ per the theoretical estimates of Ref.~\cite{Alexander:2008tq}. The variation of the annihilation signal with the assumed power-law index is mild, ranging by a factor $\sim 1.2 - 1.7$ depending on the assumed dark matter profile. 

The normalization of the power-law is set by the total number of white dwarfs within $R_0$. A recent star formation analysis, which for the first time incorporated metallicity measurements from spectroscopic surveys in this region \cite{2015ApJ...809..143D,2017MNRAS.464..194F}, estimates about $8.7 \times 10^6$ white dwarfs within the central $1.5 \, \rm pc$ \cite{2023ApJ...944...79C}. To match the observation range of Ref.~\cite{2023ApJ...944...79C}, we will only consider white dwarfs within $R \leq 1.5 \, \rm pc$ for our calculations. Fixing $R_0 = 1.5 \ \rm pc$ with the power-law index above yields a normalization $n_{\rm WD}(R_0) \simeq 3.3 \times 10^5 \ \rm pc^{-3}$. 

As our benchmarks, we consider white dwarfs with masses ranging from $0.5 - 1 \ M_\odot$, spaced by intervals of $0.1 \, M_\odot$, in order to illustrate the variation of the mass splitting sensitivity with the white dwarf profile. We do not include in our analysis white dwarfs outside this mass range, as either lighter or more massive white dwarfs require a finely-tuned progenitor evolution, and are thus generally expected to be rare, see $e.g.$ Refs.~\cite{Yoon:2007pw,2019ApJ...871..148L,2012ApJ...746...62R,2021A&A...646A..30A}. 

We derive their density and gravitational potential profiles using the Feynman-Metropolis equation of state \cite{Rotondo:2011zz,2011PhRvC..83d5805R}, assuming an admixture of carbon and oxygen in approximately equal-parts. We provide additional details on their structure in Appendix~\ref{app:WD_struc}. We will further assume these objects to have ages $\gtrsim \rm Gyr$, based on recent analyses indicating most stars in this system formed in an early star formation episode about $5 - 13 \ \rm Gyr$ ago \cite{2020A&A...641A.102S,2023ApJ...944...79C}. As detailed below, this timescale is far longer than what is largely required for their annihilation rate to equilibrate with every other process in this region. 

\subsection{IDM Density and Velocity Profile}
For the IDM density profile, we consider a benchmark Navarro-Frenk-White (NFW) profile \cite{Navarro:1995iw,Navarro:1996gj}
\begin{equation}
    \rho_\chi(R) = {\rho_\chi^0} \, {\left(\frac{R}{R_s}\right)^{-\gamma}\left(1+\frac{R}{R_s}\right)^{\gamma-3}}~,
    \label{eq:DM_halo_profile}
\end{equation}
where we fix $R_s \simeq 12 \ \rm kpc$ as the scale radius, and $\gamma$ is the slope which we allow to be in the range $1.0 - 1.5$. The standard NFW profile corresponds to $\gamma = 1$; higher slopes result in cuspier distributions which are favored by some adiabatic contraction studies \cite{2011arXiv1108.5736G,DiCintio:2014xia}. The constant $\rho_\chi^0$ is fixed so that the dark matter density at our local position is fixed to $\rho_\chi (R_\odot) = 0.42 \ \rm GeV/cm^3$, where $R_{\odot} \simeq 8.2 \ \rm kpc$ is the Galactocentric distance to the Solar System. Unfortunately, we do not find cored profiles, such as Einasto or Burkert, to contain sufficient IDM within this region for annihilation around white dwarfs to be significant. 

For the IDM velocity dispersion, for simplicity we assume a fixed value
\begin{eqnarray}
    \sigma_\chi \simeq 270 \ \rm km/s~,
\end{eqnarray}
although we note this parameter has a somewhat large uncertainty towards the galactic center. Within the region we consider, it can range $(200 - 900) \ \rm km/s$ \cite{2013PASJ...65..118S}. However, our results below only vary by an $\mathcal{O}(1)$ factor when incorporating this uncertainty, and is therefore sub-dominant relative to the main uncertainty on the total white dwarf abundance. 

\section{Kinematic Threshold}
\label{sec:cap_review} 
Since IDM is generically expected to be in its lightest state for MeV-scale mass splittings, only endothermic scattering can occur upon the particles' initial encounter with the white dwarf.
Depending on the value of the mass splitting, endothermic scattering against a given target will be kinematically allowed only within a finite region of the white dwarf interior, where the gravitational potential is sufficiently deep for the particles to reach the threshold velocity. We denote the radial distance at which the threshold is first satisfied by $r_1$. In general, this radius must be solved for numerically, based on the white dwarf's internal gravitational potential, through the condition
\begin{equation}
    w(r_1) = \sqrt{\frac{2 \delta}{m_N}}~,
    \label{eq:first_scat_rad}
\end{equation}
where $\delta$ is the mass splitting, $m_N$ is the mass of the target nucleus, and
\begin{equation}
    w(r) = \sqrt{u_\chi^2 + v^2_{\rm esc}(r)}~,
\end{equation}
is the IDM's velocity in the white dwarf's rest frame, expressed in terms of the halo velocity $u_\chi$ far from the object and the local escape velocity $v_{\rm esc}(r)$. Eq.~\eqref{eq:first_scat_rad} is valid in the non-relativistic limit, assuming a hierarchy $\delta \ll m_N \ll m_\chi$, where $m_\chi$ is the mass of the IDM's lightest state. In practice, we may neglect the initial halo velocity $u_\chi$ when computing $r_1$, based on the IDM's velocity dispersion within this region ($cf.$ Sec.~\ref{sec:gc_mod}). Note that, when $\delta$ is sufficiently small, the kinematic threshold is already met at the white dwarf's surface. We therefore cap the numerical solution to Eq.~\eqref{eq:first_scat_rad} at the white dwarf's radius, $i.e.$ we fix $r_1 = R_{\rm WD}$, in the limit $\delta \to 0$. 

In principle, each nuclear species present defines its own threshold radius $r_1$. To compute $r_1$, we fix $m_N \simeq 15\ \rm GeV$, corresponding to $^{16}$O. Choosing $^{12}$C instead rescales the threshold velocity in Eq.~\eqref{eq:first_scat_rad} by a factor $\sqrt{16/12} \simeq 1.15$, shifting $r_1$ at the percent-level given the white dwarf's escape velocity profile. Heavier elements, on the other hand, have a lower kinematic threshold and would therefore have a correspondingly larger $r_1$. In practice, however, a negligible fraction of IDM would scatter at larger radii than the value defined by oxygen. Elements with mass number beyond $^{16}$O have highly suppressed abundances and tend to sediment toward the white dwarf's center over time \cite{2022MNRAS.509.5197S, 2026ApJS..283...41B, 2021ApJ...919L..12C, 2024A&A...686A.153S}, depleting them precisely from the outer regions where a lower threshold would otherwise be relevant. Moreover, any scattering that does occur off these rarer nuclei is further suppressed by their more pronounced coherence loss (a smaller Helm form factor at fixed momentum transfer, owing to their larger nuclear radius). Therefore, from hereon we assume all of the IDM meets the threshold at $r_1$, as fixed by the oxygen mass at a given mass splitting. This shell defines a sharp boundary between the kinematically forbidden ($r>r_1$) and allowed ($r<r_1$) initial scattering regions for essentially all incoming IDM.

We additionally define the maximum mass splitting that can be excited as that corresponding to Eq.~\eqref{eq:first_scat_rad} for $r_1 = 0$, $i.e.$ when the IDM passes through the white dwarf's center at the maximum velocity it can attain,
\begin{equation}
    \delta_{\rm max} = \frac{1}{2} \, m_N \, w(0)^2~.
\end{equation}
This quantity approximately determines the maximum reach of this search in terms of inelastic mass splitting. For our white dwarf benchmarks, $\delta_{\rm max}$ roughly ranges $3 - 12 \ \rm MeV$ for inelastic scattering off oxygen.  

It is convenient to also introduce the flux of IDM particles passing through the shell $r = r_1$. This is given by
\begin{equation}
    \mathcal{F}_{\chi} = \frac{\pi r_1^2 \rho_\chi}{m_\chi} \int_{0}^{\infty} \frac{w^2(r_1)}{u_\chi} f(u_\chi) \, du_\chi~,
    \label{eq:flux_main}
\end{equation}
where
\begin{widetext}
\begin{eqnarray}
    f(u_{\chi}) = \frac{u_\chi}{v_{\rm WD}} \sqrt{\frac{3}{2\pi \sigma_\chi^2 }} \left[\exp\left(-\frac{3(u_\chi-v_{\rm WD})^2}{2 \sigma_\chi^2}\right)-\exp\left(-\frac{3(u_\chi+v_{\rm WD})^2}{2 \sigma_\chi^2}\right)\right]~
    \label{eq:relvdist}
\end{eqnarray}
\end{widetext}
is the velocity distribution in the white dwarf's frame \cite{Busoni:2017mhe}. We denote by $v_{\rm WD}$ the white dwarf's speed relative to the galactic rest frame. We fix it to $v_{\rm WD} = \sqrt{8/\pi} \, \sigma_* \simeq 285 \ \rm km \ s^{-1}$, where $\sigma_* = 179 \rm \ km \ s^{-1}$ is the inferred velocity dispersion for old stars within the Milky Way's nuclear stellar cluster \cite{Trippe:2008vj,2009A&A...502...91S}.

The IDM flux, as given by Eq.~\eqref{eq:flux_main}, accounts for the full amount of IDM passing through the region where inelastic scattering is allowed by kinematics. However, only a fraction of it will be captured and able to annihilate outside the host volume. In Secs.~\ref{sec:partial_therm} and~\ref{sec:cap-orb-ann-eq}, we estimate this fraction in terms of the IDM's particle parameters. 

\section{Partial Thermalization}
\label{sec:partial_therm}
We focus on the subset of IDM particles that, upon reaching the shell $r_1$, rapidly lose energy and fall below the inelastic scattering threshold over a small distance compared to $r_1$. This can occur either through excitation of the heavy state $\chi_2$, followed by decay, or a succession of endothermic and exothermic scatterings. 
The required number of scatters to fulfill this condition depends on the mass splitting: this parameter determines the size of the shell $r_1$, and therefore the local density, as well as the energy lost per scatter. In Sec.~\ref{sec:cap-orb-ann-eq}, we specify this quantity and relate it to the cross-section, but we note here it ranges from a single scatter in most of the white dwarf volume, to $\lesssim 40$ in the uppermost layers we are able to numerically resolve. While this assumption generally requires relatively large inelastic cross-sections, these are virtually unconstrained in the MeV-splitting regime. 

This enables us to analytically estimate the resulting orbits for the captured IDM, since the incoming particles are both captured and rapidly thermalized over a short timescale to a calculable energy and angular momentum. After slowing below threshold near the shell $r_1$, the IDM particles free-fall within the white dwarf potential. As they move inward, they are re-accelerated, and some attain velocities that again exceed the inelastic scattering threshold. We do not consider this separate population, as numerical integration of the energy loss indicates that such particles ultimately remain trapped within the white dwarf. The remaining fraction, by contrast, is unable to further lose energy, and remains pinned in long-lived orbits that extend beyond the white dwarf's volume, therefore enabling external annihilation.

The parameter separating complete from incomplete thermalization is the initial angular momentum of the IDM. We compute the critical value separating the two regimes as follows: up until inelastic scattering occurs at $r = r_1$, orbital angular momentum is conserved and given by
\begin{equation}
    J_1 = w(r_1) \, r_1  \sin\theta ~,
\end{equation}
where $\theta$ is the angle of entry into the shell of radius $r_1$. For each scattering, the deflection angle in the white dwarf rest frame is $\sim m_N/m_\chi \simeq 10^{-2}$ and therefore largely negligible for TeV-scale IDM. Similarly, decays of the up-scattered heavy state result in negligible deflections in the regime $\delta \ll m_\chi$. We therefore estimate $J_2$ by adjusting the velocity magnitude but maintaining its direction as constant,
\begin{equation}
    J_2 \simeq  c_\chi \, w_s \, r_1  \sin\theta~,
    \label{eq:J2}
\end{equation}
where $w_s = \sqrt{2 \delta/m_N}$ is the inelastic threshold velocity, and $c_\chi \lesssim 1$ is a dimensionless factor related to the number of scatters undergone by the IDM. Under the aforementioned approximation, we have kept the same radius $r_1$ for computing $J_2$.

The point of closest approach to the white dwarf's center $r_2$ is then determined from the conservation of the angular momentum $J_2$, since no further scattering proceeds. It is approximated by
\begin{equation}
    r_2 = \left(\frac{c_\chi w_s r_1}{w_2(r_2)}\right) \sin \theta~,
    \label{eq:second_scat_rad}
\end{equation}
where 
\begin{equation}
    w_2(r_2) = \sqrt{(c_\chi w_s)^2 + v^2_{\rm esc}(r_2)}~,
\end{equation}
is the IDM's velocity at the shell $r = r_2$. 

To isolate the long-lived orbits that permit external annihilation, we impose that $w_2(r_2) < w_s$, which ensures no further energy loss can proceed through inelastic scattering. This condition, combined with Eq.~\eqref{eq:second_scat_rad}, constitutes a system of equations that allows us to numerically compute a minimum angular momentum, parameterized in terms of a minimum entry angle $\theta_{\rm min}$ into the shell $r = r_1$.

\begin{figure}[t]
    \centering
    \includegraphics[width=\linewidth]{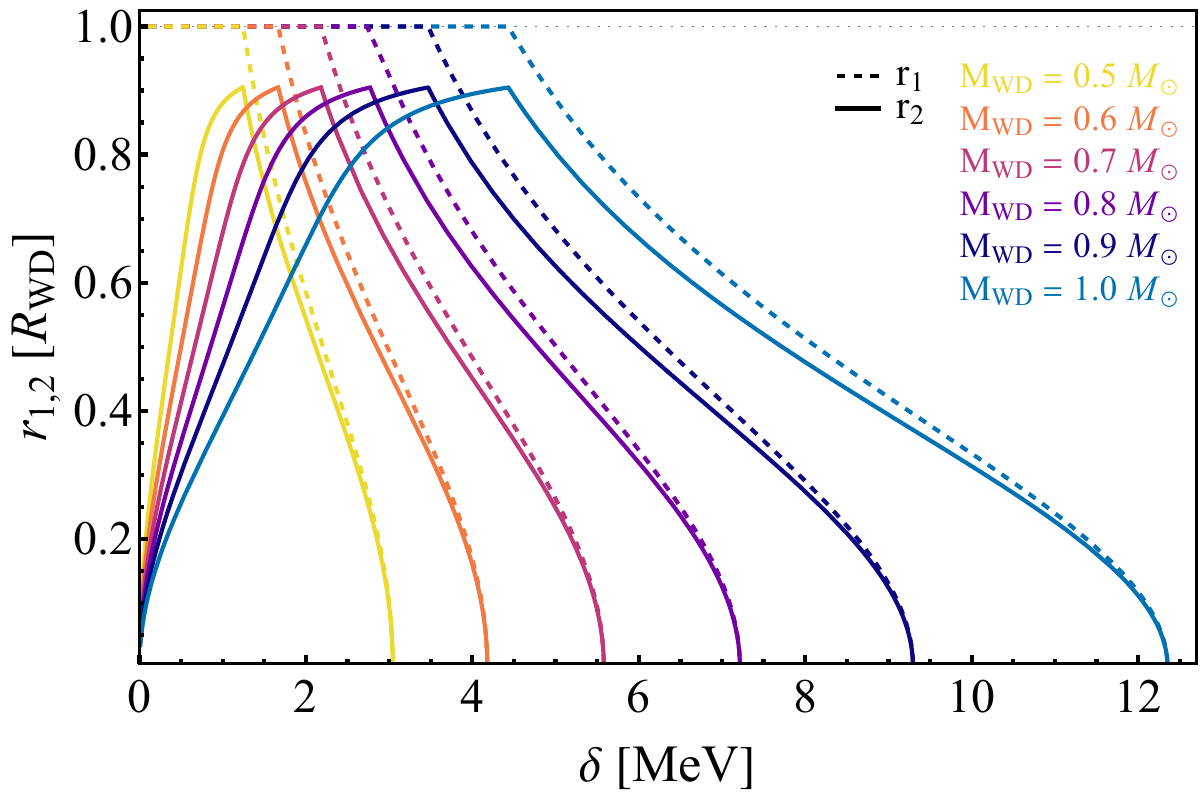}
    \caption{Capture radius $r_1$ at which inelastic scattering becomes kinematically permitted for incoming IDM, and radius of closest approach $r_2$ to the white dwarf's center after the IDM has slowed down below threshold. For the latter, we have assumed an entry angle $\theta = 0.8 \times(\pi/2)$ into the shell $r = r_1$, and a tangential velocity marginally below threshold, $i.e.$ $w(r_1) \simeq w_s$. Different colors specify the white dwarf benchmark, as labeled.}
    \label{fig:scattering_radii}
\end{figure} 
Figure~\ref{fig:scattering_radii} shows the radius $r_1$ at which the kinematic threshold is initially met, and the associated radius of closest approach $r_2$, for all our white dwarf benchmarks. When inelasticity is mild, the threshold is already met at the stellar surface, and thus $r_1 = R_{\rm WD}$. As the mass splitting increases, the threshold is met deeper within the white dwarf, displacing $r_1$ progressively inward until $r_1 = 0$ at $\delta = \delta_{\rm max}$. In this limit, only IDM particles crossing the center of the white dwarf attain a velocity large enough to up-scatter into the heavier state. The point of closest approach $r_2$, on the other hand, depends on the particle's entry angle into the shell $r=r_1$. We have fixed it to $\theta = 0.8\times(\pi/2)$ for concreteness, although qualitatively similar curves are obtained for other values. The behavior of $r_2$ reflects the resulting IDM's velocity near $r_1$ once it has slowed below the threshold. For low mass splittings, incoming IDM particles must lose a significant amount of energy before free-falling in the white dwarf's potential. As a result, $r_2 \ll r_1$ in this regime, since the particles moving very slowly near $r_1$ penetrate deep within the white dwarf. Increasing the mass splitting while $r_1 \simeq R_{\rm WD}$ translates into larger velocities near $r_1$, as a smaller amount of energy is lost by the IDM before it free-falls in the white dwarf's potential. Therefore $r_2$ tends to increase. On the other hand, once $r_1 \lesssim R_{\rm WD}$, only a small amount of energy is lost by the IDM before it is kinematically unable to scatter, and so $r_1$ and $r_2$ converge towards the same value, $i.e.$ the IDM barely penetrates beyond $r_1$ as it retains a significant amount of its energy and angular momentum. The local maximum for $r_2$ results from the transition between these two regimes. 

\begin{figure}[t]
    \centering
    \vspace*{0.4cm} 
    \includegraphics[width=\linewidth]{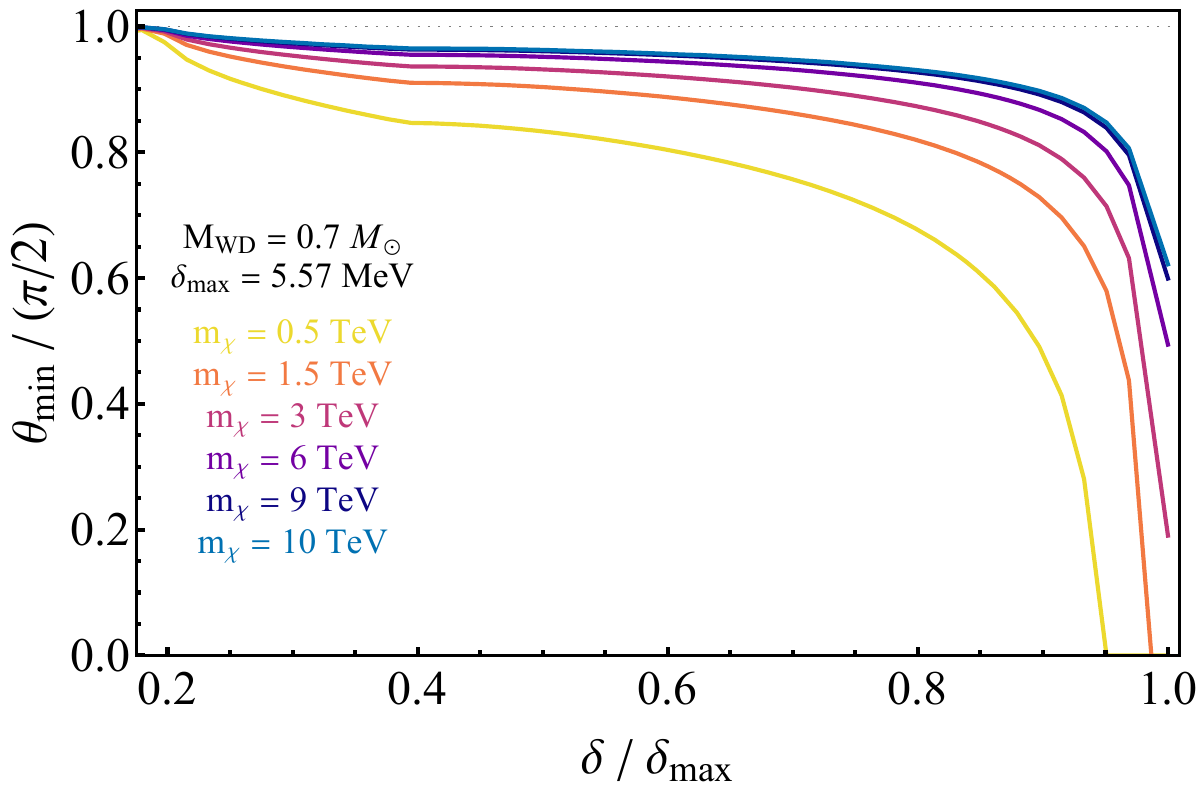}
    \caption{Minimum entry angle into the capture shell $r = r_1$ for the IDM to retain sufficient energy and angular momentum to annihilate externally, as a function of mass splitting for the specified white dwarf benchmark. Different colors indicate the assumed IDM masses, as labeled.}
    \label{fig:theta_min}
\end{figure}

Figure~\ref{fig:theta_min} shows the minimum entry angle $\theta_{\rm min}$ for which IDM particles that slowed below threshold near the shell $r=r_1$ do not scatter again at their closest approach to the center at $r=r_2$. The lines are drawn assuming a benchmark $0.7 \, M_\odot$ white dwarf and various IDM masses. For small mass splittings, $\theta_{\rm min}$ asymptotes to $\pi/2$. In this limit, inelasticity is mild, and little energy is required for the heavier state to be excited. Therefore, the IDM must reach the shell $r = r_1$ almost tangentially so that $r_2 \simeq r_1$, in order to remain below the inelastic threshold at all times. As the mass splitting increases, $\theta_{\rm min}$ decreases, since the kinematic threshold is raised. The captured IDM may therefore reach deeper layers without being up-scattered into the heavier state, permitting orbits with smaller angular momentum such that $r_2 \lesssim r_1$. 

For sufficiently light IDM, $\theta_{\rm min}$ may vanish at a finite mass splitting, indicating that the captured IDM particles are not energetic enough to re-scatter into the heavier state for any initial entry angle. By contrast, for sufficiently heavy IDM, a finite entry angle $\theta_{\rm min}$ is required throughout the full mass splitting range to allow for extended orbits, as otherwise the heavier state may be excited upon reaching deeper white dwarf layers. In this regime, $\theta_{\rm min}$ is nearly independent of the IDM mass, as the scattering kinematics are solely determined by the target's mass and the white dwarf's gravitational field. 

We emphasize that, while $\theta_{\rm min} \ll \pi/2$ might suggest that the signal is maximized since it enables a wider range of incoming angular momenta, the capture rate vanishes as $\delta \to \delta_{\rm max}$. This is because only IDM particles with nearly zero angular momentum passing through the center meet the kinematic threshold, and these constitute a tiny flux. Consequently, the annihilation signal also vanishes in this limit. 

Finally, we note that, so far, we have not considered loop-level elastic scattering, which is generically expected to be present and does not impose a threshold energy. While we assume the elastic scattering channel is suppressed relative to its inelastic counterpart, its long-term effect is to induce orbital decays since its associated energy loss rate, though significantly smaller than that during initial thermalization, remains finite. Below, we compare this orbital decay process against capture and annihilation.

\section{Capture-Annihilation-Orb.~Decay Equilibrium}
\label{sec:cap-orb-ann-eq}
Let $N_\chi$ be the number of particles with orbits \textit{extending beyond} the white dwarf's volume. This quantity evolves over time as 
\begin{equation}
    \frac{dN_\chi}{dt} = \Gamma_{\rm cap} - \Gamma_{\rm orb} \, N_\chi - \Gamma_{\rm ann} \, N_\chi^2
    \label{eq:dNdt}
\end{equation}
where the first term corresponds to the rate at which particles are captured onto such orbits, the second term is the rate at which those orbits decay, and the third term is the rate of depletion due to annihilation. For a given set of coefficients $\Gamma_{\rm cap}$, $\Gamma_{\rm orb}$ and $\Gamma_{\rm ann}$, the solution starting from the initial condition $N_\chi (t = 0) = 0$ is 
\begin{widetext}
\begin{eqnarray}
      N_\chi (t) = \frac{\sqrt{4 \, \Gamma_{\rm ann} \,\Gamma_{\rm cap} + \Gamma^2_{\rm orb}}}{2 \Gamma_{\rm ann}} \,\tanh{\left[{\frac{1}{2}\sqrt{4 \, \Gamma_{\rm ann} \, \Gamma_{\rm cap} + \Gamma^2_{\rm orb} } \, \, t} + \tanh^{-1}\left({\frac{\Gamma_{\rm orb}}{\sqrt{ 4 \, \Gamma_{\rm ann} \, \Gamma_{\rm cap} + \Gamma^2_{\rm orb}}}}\right)\right]}  - \frac{\Gamma_{\rm orb}}{2 \Gamma_{\rm ann}} ~.
      \label{eq:Nchi_full}
\end{eqnarray}
\end{widetext}

This introduces the equilibration timescale
\begin{eqnarray}
    \tau_{\rm eq} = \frac{2}{\sqrt{4 \, \Gamma_{\rm ann} \, \Gamma_{\rm cap}+\Gamma^2_{\rm orb} }}~,
    \label{eq:eq_timescale}
\end{eqnarray}
such that, when $t \gtrsim \tau_{\rm eq}$, the number of particles asymptotes to the equilibrium value 
\begin{eqnarray}
     N^{\rm eq}_\chi \simeq \frac{1}{2 \, \Gamma_{\rm ann}}\left( \sqrt{4 \, \Gamma_{\rm ann} \, \Gamma_{\rm cap} + \Gamma_{\rm orb}^2 } - {\Gamma_{\rm orb}}\right)~.
     \label{eq:eq_Nchi}
\end{eqnarray}
When this regime is reached, the annihilation rate attained by an individual white dwarf is then given by the last term of Eq.~\eqref{eq:dNdt}, evaluated at the equilibrium number,
\begin{align}
     \label{eq:ann_rate_single_WD}
    \mathcal{A}_{\rm eq} & = \frac{1}{2} \, \Gamma_{\rm ann} \left(N_{\chi}^{\rm eq}\right)^2 \\ & = \frac{1}{8 \, \Gamma_{\rm ann}} \left( \sqrt{4 \, \Gamma_{\rm ann} \,\Gamma_{\rm cap} + \Gamma^2_{\rm orb} }-\Gamma_{\rm orb}\right)^2~. \nonumber
\end{align}
Relative to Eq.~\eqref{eq:dNdt}, an additional factor $1/2$ is introduced to account for the depletion of two particles per annihilation event.

As can be seen from the above expressions, the regime where orbital decays are negligible relative to capture and annihilation corresponds to
\begin{eqnarray}
    \Gamma^2_{\rm orb} \ll 4\,\Gamma_{\rm ann}\,\Gamma_{\rm cap}~.
\end{eqnarray}
In this limit, the equilibration timescale, particle number, and annihilation rate converge to their usual formulae
\begin{eqnarray}
    \tau_{\rm eq} \to {1} \,/\,{\sqrt{\Gamma_{\rm cap} \, \Gamma_{\rm ann}}} ~,
\end{eqnarray}
\begin{eqnarray}
     N^{\rm eq}_\chi \to \sqrt{\Gamma_{\rm cap}\, / \,\Gamma_{\rm ann}}~,
\end{eqnarray}
\begin{eqnarray}
    \mathcal{A}_{\rm eq} \to \Gamma_{\rm cap} \,/\,2~.
\end{eqnarray}
Comparing the above to Eqs.~\eqref{eq:eq_timescale}$-$\eqref{eq:ann_rate_single_WD}, the effect of orbital decays is to suppress the equilibrium number of IDM particles. The equilibration timescale is subsequently shorter, since a smaller number of particles must be accumulated before capture balances both depletion mechanisms. Because the equilibrium particle number is reduced in the presence of orbital decays, the resulting annihilation rate is also suppressed compared to the regime where they are negligible. 
It is worth remarking that, although the standard formulae are recovered when orbital decay is negligible, the equilibrium values will be modified compared to prior analyses. This is because the coefficients only account for a subset of all incoming IDM particles. For instance, the capture rate $\Gamma_{\rm cap}$ applies the entry angle cut, and is therefore smaller than the full capture rate of the object.

Each coefficient appearing in Eq.~\eqref{eq:dNdt} is an implicit function of the white dwarf properties, as well as the various particle parameters of the IDM. Below, we proceed to compute approximate expressions for each of these, valid within the framework outlined in the previous sections. 

\subsection{Capture Coefficient}
The capture coefficient $\Gamma_{\rm cap}$ is determined from the flux given by Eq.~\eqref{eq:flux_main}, corrected to only account for the incoming IDM particles that 
enter the shell $r_1$ with an entry angle $\theta \geq \theta_{\rm min}$ and also rapidly slow down below threshold. To introduce these corrections, we first recast the flux given by Eq.~\eqref{eq:flux_main} in terms of angular momentum, and introduce a probability factor to account for the required number of scatters,
\begin{align}
    \label{eq:gamma_cap_gen}
    \Gamma_{\rm cap} = \frac{\pi \rho_\chi}{m_\chi} \int_{J_{\rm min}}^{J_{\rm max}} P(k \geq k_{\rm min}) \, dJ^2 \\\times \int_0^{\infty} & \frac{f(u_\chi)}{u_\chi}  \, du_\chi~, \nonumber
\end{align}
where the integration limits are 
\begin{eqnarray}
    J_{\rm min} = r_1 \, w(r_1) \sin \theta_{\rm min}~,
\end{eqnarray}
\begin{eqnarray}
    J_{\rm max} = r_1 \, w(r_1)~.
\end{eqnarray}

We denote by $P(k \geq k_{\rm min})$ the probability to attain a number of scatters $k$ above some minimum $k_{\rm min}$ required to slow below threshold. We assume this is given by a Poisson distribution, which in terms of the cumulative distribution function reads
\begin{eqnarray}
    P(k \geq k_{\rm min}) = 1 - \mathrm{CDF}(\mathrm{Pois}(\tau), \, k_{\rm min})~.
    \label{eq:opt_thick_2}
\end{eqnarray}
The expectation $\tau$ is the optical depth for inelastic scattering, defined as 
\begin{eqnarray}
    \tau(\theta) = \int_0^{2 \, r_1 \cos\theta} n_{N}(r) \, \sigma^{(N)}_{\rm Inel} \, dl ~,
\end{eqnarray}
where $n_N$ is the number density of target nuclei, $\sigma^{(N)}_{\rm Inel}$ is the nuclear-level inelastic cross-section. For the range of velocities and mass splittings of interest, we relate it to the nucleon-level cross-section $\sigma^{(n)}_{\rm Inel}$ via $\sigma^{(N)}_{\rm Inel} \simeq 0.01 \, A^4 \, \sigma^{(n)}_{\rm Inel}$, where the numerical factor approximately accounts for the partial loss of coherence from the Helm form factor at the momentum transfers involved throughout capture and thermalization. We have also approximated the IDM's trajectory across the shell as linear, with $r^2 = l^2 + r_1^2 - 2\,l\,r_1\cos\theta$. We compute the minimum number of inelastic scatters as 
\begin{eqnarray}
    k_{\rm min} \simeq {\rm Max}\left[1, \,\frac{1}{2 \delta} \times \frac{1}{2} \, m_\chi \left(w(r_1)^2 - w_s^2\right)\right]~,
\end{eqnarray}
where the energy lost per scatter is approximated by the factor $2\delta$ in the denominator: the recoil energy of the nucleus is of order $\sim \delta$, while the second factor of $\delta$ derives from the inelastic transition itself.

The probability factor $P(k \geq k_{\rm min})$ effectively rejects cases in which the IDM reaches the shell $r_1$ but does not scatter enough times to slow below the inelastic threshold. In practice, this largely becomes relevant at small mass splittings, for which the energy loss per scatter is tiny and scattering occurs near the white dwarf surface, where the target density is lowest. This is also the regime in which our linear-trajectory approximation is most justified: since the entry angle $\theta_{\rm min}$ must be close to $\pi/2$ for the IDM to externally annihilate ($cf.$ Fig.~\ref{fig:scattering_radii}), the corresponding path lengths are short, and neglecting the slight curvature of the IDM's trajectory across the white dwarf is reasonable. By contrast, at larger mass splittings, $r_1$ shifts toward the white dwarf's inner core, and a single scatter suffices in cases where the outgoing heavy state promptly decays (or at most one endothermic scattering followed by an exothermic one, if the decay is not sufficiently fast). In this regime, $P(k \geq k_{\rm min}) \simeq 1$ for substantially smaller inelastic cross-sections.

Note that, for each collision, we assume the target nucleus to be at rest. Nuclei in old white dwarfs that have undergone crystallization have velocities determined by the in-medium ion plasma frequency. However, this velocity is relatively negligible compared to that of the IDM throughout the thermalization process. The potential breakdown of this approximation occurs in the innermost layers of the heaviest white dwarf benchmark, where the density and therefore the ion plasma frequency reaches significant values. However, IDM that initially scatters at such depths undergoes rapid orbital decay, and therefore does not contribute to the signal computed below. See Ref.~\cite{Acevedo:2023xnu} for further details on the impact of ion motion on dark matter capture in various kinematic regimes. 

\subsection{Orbital Decay Coefficient}
The orbital decay of the captured IDM is ultimately caused by elastic scatterings, which are generically expected to proceed at loop-level. A naive estimate indicates multiple elastic scatters are required for the orbit to fully contract beneath the stellar surface. However, we find numerically that even a single elastic scatter can alter the IDM's orbit so that it accesses deeper white dwarf layers. When this occurs, the threshold for inelastic scattering can be met again, initiating rapid energy loss. Therefore, the timescale for the IDM to elastically scatter once is approximately the bottleneck of the orbital decay process. In general, this depends on the exact details of the IDM's trajectory. Particles that initially had entry angles into the capture shell close to $\theta_{\rm min}$ will tend to elastically scatter first, as these periodically transit denser regions of the white dwarf, and vice versa. 

For a given entry angle $\theta$, we approximate the average time for the IDM to elastically scatter once as 
\begin{eqnarray}
    \mathcal{T}(\theta) \simeq {P_{r_2 \rightarrow R_{\rm WD}}} \left( \int_{r_2}^{R_{\rm WD}} n_{N}(r) \, \sigma^{(N)}_{\rm Elas} \, dr \right)^{-1},
    \label{eq:time_elastic}
\end{eqnarray}
where $\sigma^{(N)}_{\rm Elas}$ is the nuclear-level total elastic cross-section. It is related to the nucleon-level elastic cross-section via $\sigma^{(N)}_{\rm Elas} \simeq (0.03-0.1) \, A^4 \,\sigma^{(n)}_{\rm Elas}$, where $\sigma^{(n)}_{\rm Elas}$ is the nucleon-level cross-section, $A$ is the nuclear mass number, and the factor in brackets approximately spans the partial loss of coherence over the orbital velocities of interest. For the latter, the lower bound is a more accurate representation of the true value, since the probability for elastic scattering is substantially larger in the denser regions, despite the correspondingly larger coherence loss there. The factor on the right,
\begin{eqnarray}
    P_{r_2 \rightarrow R_{\rm WD}} = \int^{R_{\rm WD}}_{r_2} \frac{dr}{|v_r|}~,
    \label{eq:radial_period}
\end{eqnarray}
is the time it takes for the particle to transit from its orbital pericenter $r_2$ to the white dwarf's surface, expressed in terms of the radial velocity
\begin{eqnarray}
    |v_r| = \sqrt{2 (\epsilon - \phi(r)) - J^2/r^2}~.
    \label{eq:rad_velocity}
\end{eqnarray}
Above, $\phi(r) = -v^2_{\rm esc}(r)/2$ is the white dwarf's gravitational potential, and $\epsilon$ is the IDM's energy per unit mass. The dependence on the entry angle $\theta$ is implicit through the distance of closest approach $r_2$ and the angular momentum $J$. It should be noted that Eq.~\eqref{eq:time_elastic} is approximate, since the numerator neglects the curvature of the IDM's trajectory, although expect this to introduce a correction of order unity.

Formally, the orbital decay coefficient is 
\begin{eqnarray}
    \Gamma_{\rm orb} = \frac{1}{N_\chi} \, {\int \frac{1}{\mathcal{T}(\theta)} \, \frac{dN_\chi}{d\theta} \, d\theta}~,
    \label{eq:gamma_orb}
\end{eqnarray}
\begin{eqnarray}
    N_\chi = \int \frac{dN_\chi}{d\theta} \, d\theta~.
\end{eqnarray}
This is expressed in terms of the differential occupation number per unit entry angle
\begin{eqnarray}
    \frac{dN_\chi}{d \theta} = \mathcal{T}(\theta) \times \frac{d \Gamma_{\rm cap}}{d\theta}~,
    \label{eq:dNchidtheta}
\end{eqnarray}
where
\begin{eqnarray}
    \frac{d{\Gamma_{\rm cap}}}{d\theta} = 2 \cos\theta \sin\theta \, \,\frac{d\Gamma_{\rm cap}}{d(J/ J_{\rm max})^2}
\end{eqnarray}
is the differential rate at which the orbit bundle spanned by $d\theta$ is populated through IDM capture, $cf.$ Eq.~\eqref{eq:gamma_cap_gen}. We emphasize that integrating the differential occupation number as defined above does not yield the physical number of IDM particles; this is set by the competition between all intervening processes as per Eq.~\eqref{eq:Nchi_full}. Rather, the above must be interpreted as a weight factor for each set of orbits spanned by $d\theta$, determined by the ratio between particle depletion and replenishment rates.

We remark that $\Gamma_{\rm orb}$ depends on the elastic cross-section through the timescale $\mathcal{T}(\theta)$, which also enters the denominator in Eq.~\eqref{eq:gamma_orb}. This parameter largely determines whether or not orbital decays dominate as the main particle depletion mechanism. The orbital decay coefficient is also, as expected, independent of the background density $\rho_\chi$ surrounding the white dwarf. This is because orbital decay is driven solely by elastic scattering processes, irrespective of the total amount of IDM present in the environment. The dependence on $\rho_\chi$ cancels between the numerator and denominator of Eq.~\eqref{eq:gamma_orb}, since both integrands are proportional to this quantity. This, in particular, indicates that orbital decay generally becomes less significant in regions of high dark matter density, like the Galactic Center. In these environments, particles depleted through orbital decay are rapidly replenished by newly captured particles owing to the enhanced capture rate.

\subsection{Annihilation Coefficient}
Finally, the annihilation coefficient is given by
\begin{equation}
    \Gamma_{\rm ann} = \frac{1}{N^2_\chi} \, {\int \langle \sigma v \rangle\, n^2_\chi(r) \, dV} ~,
    \label{eq:ann_coeff_basic}
\end{equation}
where $\langle \sigma v \rangle$ is the annihilation cross-section, and $n_\chi(r)$ is the number density profile of IDM particles able to undergo external annihilation. In contrast with the orbital decay coefficient, computations are more convenient if framed in terms of $n_\chi(r)$ rather than $dN_\chi/d\theta$. The reason is that particles with different entry angles may annihilate wherever their trajectories overlap. Consequently, the numerator in $\Gamma_{\rm ann}$, if written in terms of $dN_\chi/d\theta$, leads to a more complicated double integral over the independent entry angles of both annihilating particles into the capture shell $r = r_1$, with additional cuts that exclude the regions where the particles' trajectories do not overlap. 

We express the density profile as
\begin{eqnarray}
    n_{\chi}(r) = \int^{\pi/2}_{\theta_{\rm min}}  \frac{dn_\chi}{ d\theta} \, \Theta( r_T(\theta)- r) \, d\theta~,
    \label{eq:chi_profile}
\end{eqnarray}
where the Heaviside function $\Theta$ rejects any contribution from particles with entry angles such that their radial turning points 
\begin{equation}
     r_T(\theta) \simeq \frac{G M_{\rm WD}}{2 |\epsilon|} \left[1 + \sqrt{1 + 2 \epsilon \left(\frac{J}{GM_{\rm WD}}\right)^2}\right]~,
\end{equation}
are smaller than $r$. In other words, it filters out particles lacking the angular momentum to reach the radius at which the density is evaluated. Note that this exterior turning point can be obtained analytically, since the gravitational potential simply scales $\propto 1/r$ outside the white dwarf's volume.

We calculate differential density per unit entry angle as 
\begin{equation}
    \frac{dn_\chi}{d\theta} = \frac{d^2N_\chi}{dV d\theta} =  \frac{1}{4 \pi r^2} \, \frac{d^2N_\chi}{dr \, d\theta}~,
\end{equation}
where we further decompose
\begin{eqnarray}
    \frac{d^2N_\chi}{dr\,d\theta} = \frac{dN_\chi}{d\theta} \times \frac{dp}{dr}~.
    \label{eq:dNchi_drdtheta}
\end{eqnarray}
The first factor is given by Eq.~\eqref{eq:dNchidtheta}, whereas $dp/dr$ is the probability of encountering a particle at a shell $r$ with width $dr$. We approximate the latter as the ratio of the time spent at such volume relative to its full orbital period. In particular, the time spent within a small volume centered at $r$, with characteristic length $dr$, is given by the ratio of $dr$ to the velocity at that location. Therefore, we have
\begin{eqnarray}
    \frac{dp}{dr} = \frac{1}{|v_r|} \times \frac{1}{P_{r_2 \rightarrow r_{T}}}~,
\end{eqnarray}
where we now use the full radial period 
\begin{eqnarray}
    P_{r_2 \rightarrow r_T} = P_{r_2 \rightarrow R_{\rm WD}}  + \int^{r_T}_{R_{\rm WD}} \frac{dr}{|v_r|} \,~,
    \label{eq:radial_period_full}
\end{eqnarray}
$cf.$ Eq.~\eqref{eq:radial_period}. The second term in Eq.~\eqref{eq:radial_period_full} can also be obtained exactly, owing to the simple analytic form of the gravitational potential outside the star, 
\begin{widetext}
\begin{eqnarray}
    \int^{r_T(\theta)}_{R_{\rm WD}} \frac{dr}{|v_r|} =  \left[ -\frac{\sqrt{- 2 |\epsilon| r^2 + 2 G M_{\rm WD} r - J^2}}{2 |\epsilon|} + \frac{GM_{\rm WD}}{(2|\epsilon|)^{3/2}} \, {\rm sin}^{-1} \left(\frac{2 |\epsilon| r - GM_{\rm WD}}{\sqrt{(GM_{\rm WD})^2 - 2 |\epsilon| J^2}}\right)\right]_{R_{\rm WD}}^{r_T} ~.
\end{eqnarray}
\end{widetext}

As with the orbital decay coefficient, the integration to obtain the IDM profile $n_\chi(r)$, and therefore the annihilation coefficient itself, must ultimately be performed numerically, since no closed-form expression is available for the white dwarf's internal potential profile. In what follows, we limit ourselves to a fixed, velocity-independent (s-wave) annihilation cross-section for simplicity. However, it should be noted that the generalization to a velocity-dependent cross-section is straightforward. The additional velocity dependence can be parameterized as a function of radius, and absorbed into the volume integral in Eq.~\eqref{eq:ann_coeff_basic}. 

\subsection{Comparison between Coefficients}
Having computed each coefficient above, it is instructive to compare them in order to illustrate the parameter range for which a strong annihilation signal can be attained. Based on the equilibrium annihilation rate, Eq.~\eqref{eq:ann_rate_single_WD}, we focus on the combination 
\begin{eqnarray}
    \sqrt{4 \, \Gamma_{\rm ann} \,\Gamma_{\rm cap} + \Gamma^2_{\rm orb}} - \Gamma_{\rm orb}~.
    \label{eq:coeff_comparison}
\end{eqnarray}
This quantity determines whether orbital decay dominates over annihilation as the primary depletion channel for the IDM particles orbiting around the white dwarf.

\begin{figure}[t!]
    \centering
    \includegraphics[width=\linewidth]{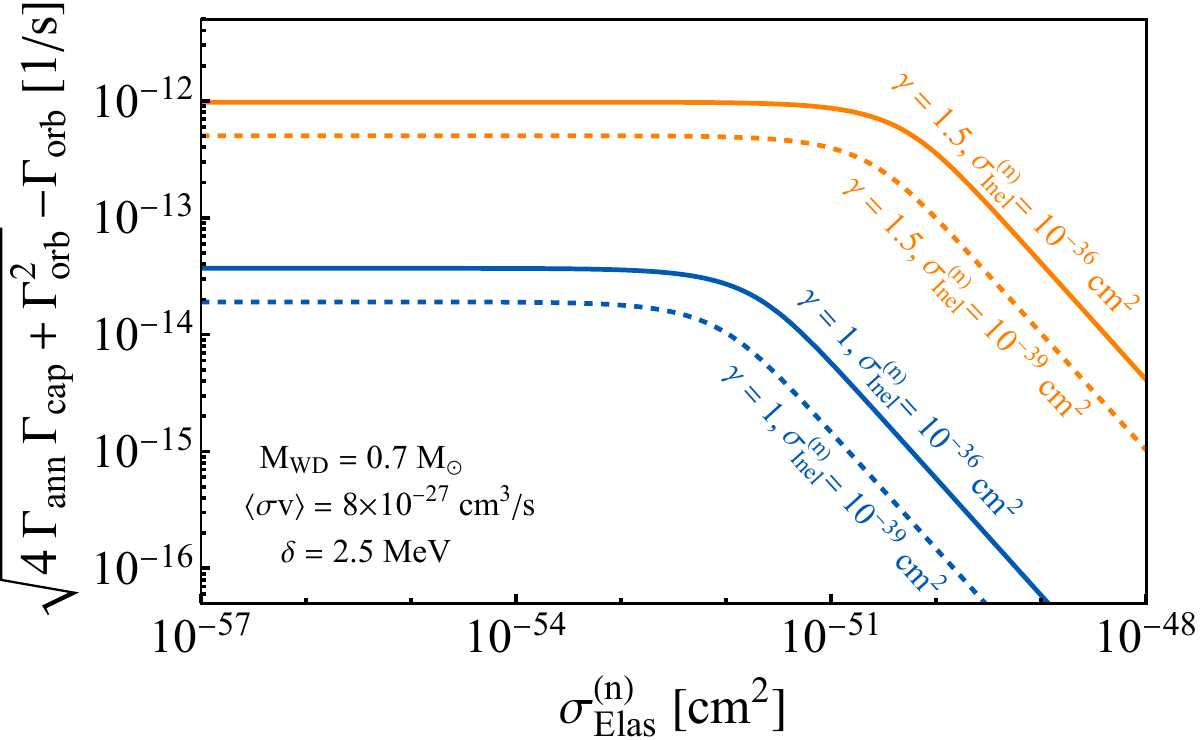}
    \caption{Combination of capture, orbital decay, and annihilation coefficients that characterizes the equilibrium regime among these processes ($cf.$ Eqs.~\eqref{eq:eq_Nchi} and \eqref{eq:ann_rate_single_WD}), as a function of the elastic nucleon scattering cross-section. We have fixed the white dwarf benchmark, mass splitting, and annihilation cross-section, as specified. Colors denote the assumed inner slope of the IDM density profile toward the Galactic Center, while solid and dashed lines correspond to different fiducial values of the inelastic nucleon cross-section.}
    \label{fig:coeff_comparison}
\end{figure}

Figure~\ref{fig:coeff_comparison} shows~\eqref{eq:coeff_comparison} as a function of elastic cross-section, for a benchmark $\sim 0.7 \, M_\odot$ white dwarf at a distance $0.01 \ \rm pc$ from the Galactic Center (corresponding to the small-distance cutoff we consider, see below). We have fixed a fiducial annihilation cross-section $\langle \sigma v\rangle = 8 \times 10^{-27} \, \rm cm^3/s$, and mass splitting $\delta = 2.5 \ \rm MeV$. Each set of curves corresponds to a different assumed IDM halo density slope, for two different nucleon-level inelastic cross-sections. In each case, small elastic cross-sections render orbital decays a negligible process, and annihilation equilibrates with capture. As the elastic cross-section increases, orbital decays proceed at a faster rate, and eventually this process overtakes annihilation as the depletion mechanism for the captured IDM. In this regime, the annihilation rate, as given by Eq.~\eqref{eq:ann_rate_single_WD}, becomes suppressed. The exact transition point depends on the background IDM density, as well as the inelastic and annihilation cross-sections: these parameters regulate capture and annihilation rates, and therefore determine the elastic cross-section required for orbital decay to compete with these processes. While not explicitly shown, increasing the mass splitting has the dual effect of making orbital decays stronger, since the resulting orbits traverse deeper regions of the white dwarf, while at the same time boosting the annihilation rate, since particles with a wider range of angular momenta now contribute to annihilation. The variation of the inelastic cross-section reach with this parameter is shown further below. 

Finally, we comment on the equilibration timescale $\tau_{\rm eq}$, $cf.$ Eq.~\eqref{eq:eq_timescale}. While its exact value depends on the assumed model parameters and background IDM density, we find this timescale to be $\ll 1 \ \rm Gyr$ ($\simeq 3.15 \times 10^{16} \ \rm s$) throughout our study. This is, in particular, already illustrated by Fig.~\ref{fig:coeff_comparison}: when orbital decays are negligible, the equilibration timescale is the inverse asymptote at small elastic cross-sections. Conversely, when orbital decays are dominant, the equilibration timescale is even shorter (though no longer given by the inverse lines in Fig.~\ref{fig:coeff_comparison}). In this case, orbital decay balances with capture at an earlier time after a relatively small particle population is attained. Therefore, in what follows we shall consider equilibration among all three processes to be established, and the annihilation rate to be accordingly given by Eq.~\eqref{eq:ann_rate_single_WD}.

\section{Annihilation Signal}
\label{sec:collective_signal}
\subsection{Integrated Rate}
We now turn to the integrated signal from the Galactic Center. As the stellar and IDM distributions within our region of interest are spherically symmetric to a good approximation, the collective annihilation rate is given by the volume integral
\begin{equation}
    \mathcal{A}_{\rm WDs} = \int^{R_{\rm max}}_{R_{\rm min}} \mathcal{A}_{\rm eq} \, n_{\rm WD}(R) \, 4 \pi R^2 \, dR   ~.
    \label{eq:tot_ext_ann}
\end{equation}
As shown above, the equilibration time is short compared to the range of expected white dwarf ages in this region. Therefore, we have fixed the annihilation rate per white dwarf as $\mathcal{A}_{\rm eq}$, $cf.$ Eq.~\eqref{eq:ann_rate_single_WD}. The above is integrated from a minimum $R_{\rm min} = 10^{-2} \ \rm pc$, up to a value $R_{\rm max} =  1.5 \ \rm pc$, where the upper cutoff is taken in order to remain within the data range of Ref.~\cite{2023ApJ...944...79C}, which we have used to normalize our white dwarf distribution, $cf.$ Eq.~\eqref{eq:wd_dist}. The lower integration limit, on the other hand, is chosen so as to avoid the largest uncertainties in the IDM density and velocity dispersion closer to Sgr.~A$^{\star}$. 

Eq.~\eqref{eq:tot_ext_ann} assumes a single-mass spectrum for the white dwarf population. The generalization over a spectrum of masses is straightforward. Assuming for now a $\sim 0.7 \, M_\odot$ monochromatic mass spectrum, and negligible orbital decays, the maximum estimated collective rate is 
\begin{align}
    \mathcal{A}_{\rm WDs} \simeq (4 \times 10^{34} - 1.1 \times 10^{37}) \, \, \rm s^{-1} ~,
    \label{eq:full_wd_rate_num}
\end{align} 
for TeV IDM, where we have bracketed the full range of IDM profile slopes $\gamma = 1 - 1.5$. We can compare the above to its halo counterpart, given by 
\begin{equation}
    \mathcal{A}_{\rm halo} = \int \left(\frac{\rho_\chi(R)}{m_\chi}\right)^2 \langle \sigma v\rangle \, 4 \pi R^2 dR ~.
\end{equation}
We evaluate it over the volume spanned by the analysis region of H.E.S.S. that we consider in this work, set roughly by the $\sim 0.13^\circ$ angular resolution of this instrument \cite{Malyshev:2015hqa}. In terms of galactocentric distance, this corresponds to $\sim 18 \ \rm pc$. Assuming a fixed, velocity-independent annihilation cross-section, the above expression evaluates to
\begin{align}
  \label{eq:halo_ann_detailed}
    \mathcal{A}_{\rm halo}  \simeq (3.5 \times 10^{33} - 1.8 \times 10^{37}) \, \, \rm s^{-1} \\ \times \left(\frac{\rm TeV}{m_\chi}\right)^2 & \left(\frac{\langle \sigma v \rangle}{8 \times 10^{-27} \ \rm cm^3/s}\right)~, \nonumber
\end{align} 
where we have bracketed the uncertainty spanned by the DM profile slope range $\gamma = 1 - 1.5$, $cf.$ Eq.~\eqref{eq:DM_halo_profile}.

The above illustrates that there is a range of parameters for which the signal sourced by white dwarfs within the central parsec dominates over the halo signal of a much wider region. In particular, we find this to be the case when the effect of orbital decays is negligible to mild. We note, however, that for the steepest profile slope we consider, the white dwarf-sourced and halo annihilation rates can become comparable, though this ultimately depends on the specific relation between the annihilation and elastic cross-section. For instance, lowering both would keep orbital decays negligible, and therefore Eq.~\eqref{eq:full_wd_rate_num} would remain unchanged, whereas the halo rate would correspondingly drop. 
It is also worth noting that the enhancement to the annihilation rate from white dwarfs would be even greater in scenarios where the cross-section is velocity-suppressed, such as p-wave annihilation. This is because the characteristic speed of captured IDM around white dwarfs is substantially higher compared to its ambient velocity dispersion, of order $(1-3) \times 10^{-2} \,$ compared to $\sigma_\chi \sim 10^{-3} \,$. 

\subsection{Flux}
In general, the resulting gamma-ray spectral flux is 
\begin{equation}
    \frac{d\Phi_{\gamma}}{dE_\gamma} = \frac{\mathcal{A}_{\rm WDs}}{4\pi D_{\rm GC}^2} \sum_{\rm SM} {\rm Br}({\chi \bar{\chi} \rightarrow \rm SM}) \, \frac{dN_{{\gamma}}}{dE_{\gamma}}\bigg|_{\rm SM}  ~.
\label{eq:gamma_flux}
\end{equation}
Above, $E_\gamma$ is the photon energy, Br$\left(\chi \bar{\chi} \rightarrow {\rm SM}\right)$ is the annihilation branching ratio for a given channel denoted by $\rm SM$, $dN_{\gamma}/dE_\gamma |_{\rm SM}$ denotes the gamma-ray spectrum per annihilation for that channel, and $D_{\rm GC} \simeq 8.2 \ \rm kpc$ is the distance from the observation point to the Galactic Center. In general, gamma-rays can undergo attenuation as they propagate through the Milky Way's interstellar medium. However, given the IDM mass range we consider, it is always the case that either primary or secondary photons have energies $E_\gamma \lesssim 1 \ \rm TeV$, for which attenuation effects are mild to negligible \cite{Fang:2021ylv}.
\begin{figure*}[t!]
    \centering
     \includegraphics[width=0.325\textwidth]{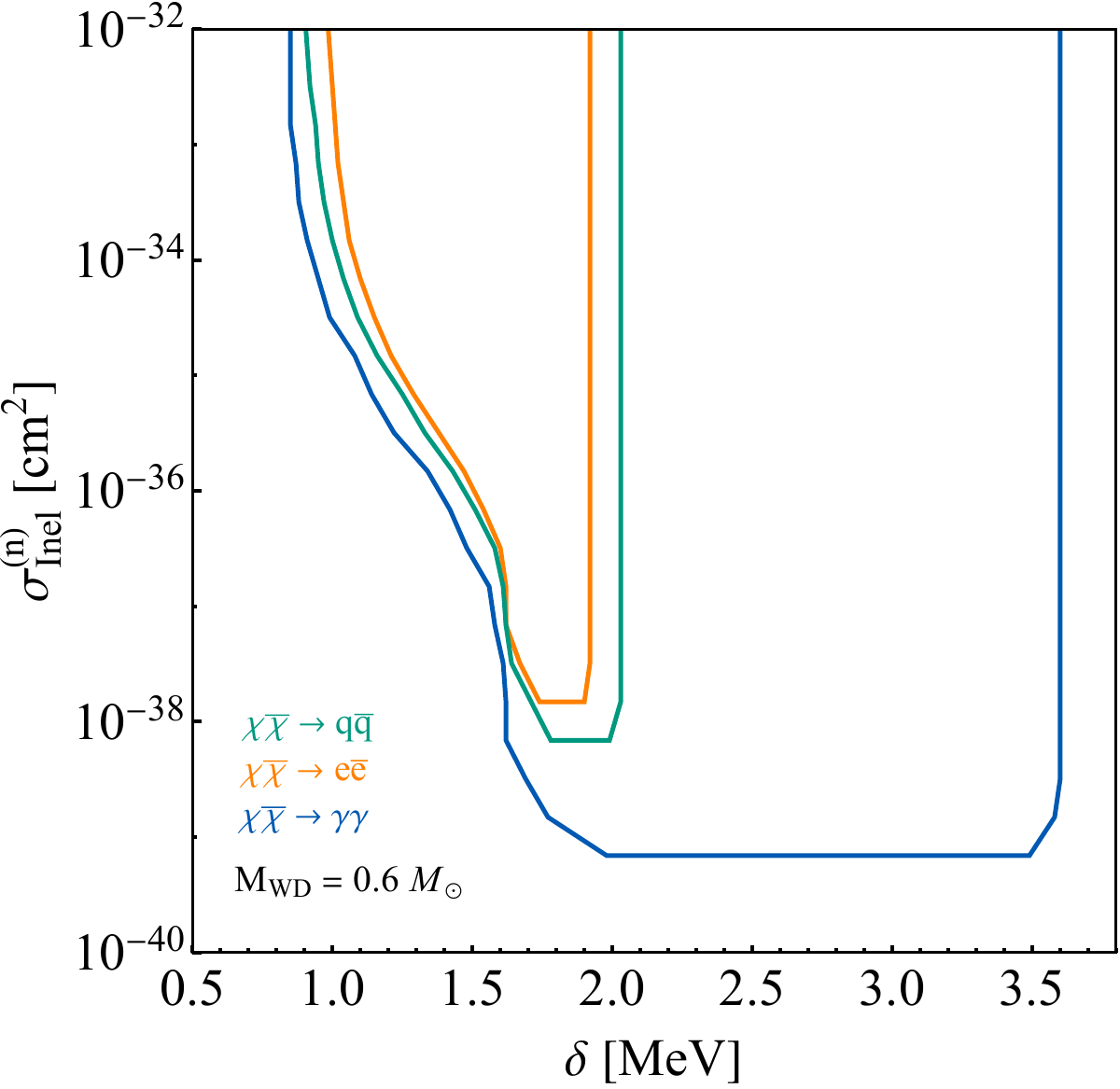}
     \includegraphics[width=0.325\textwidth]{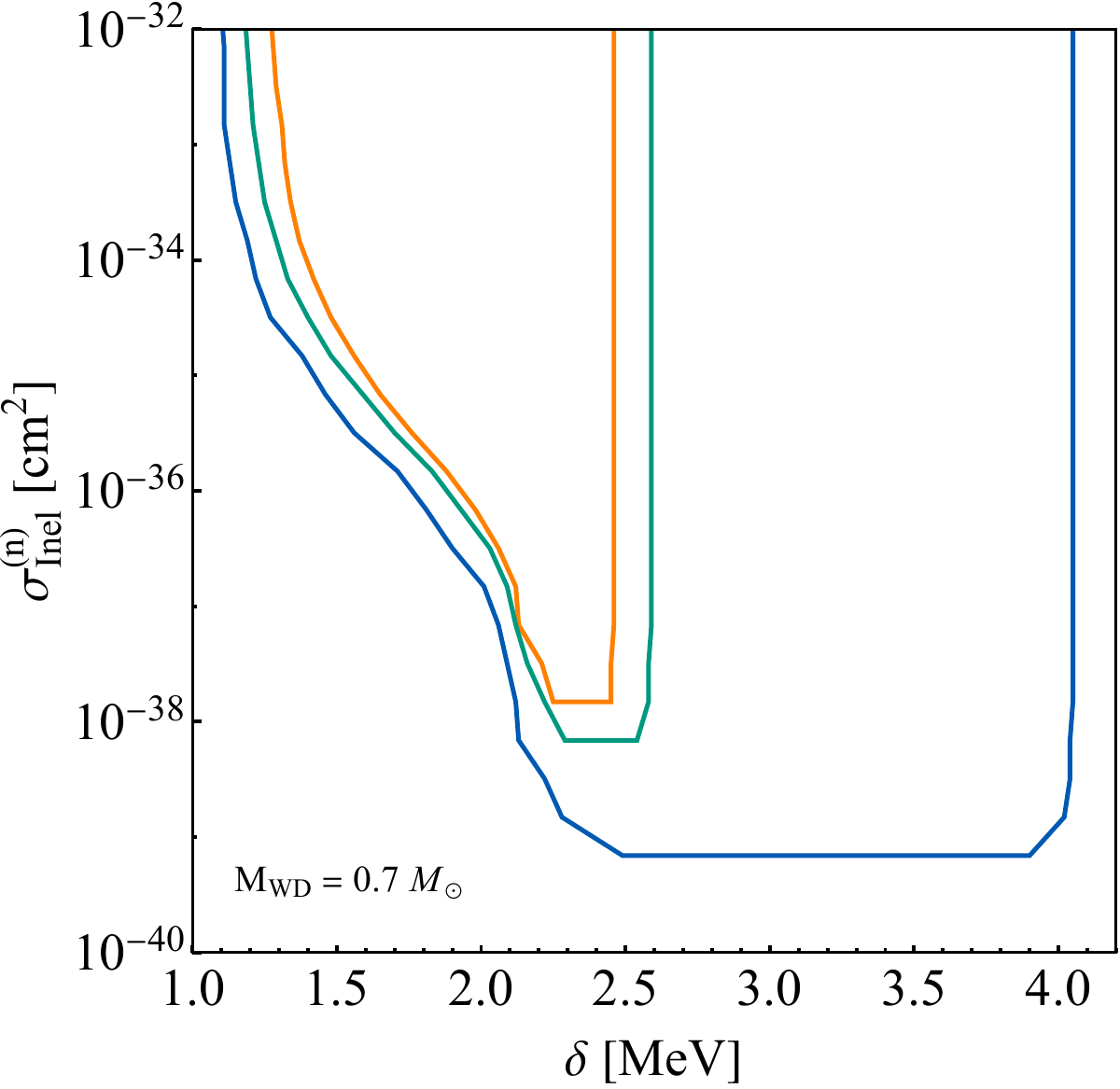}
     \includegraphics[width=0.325\textwidth]{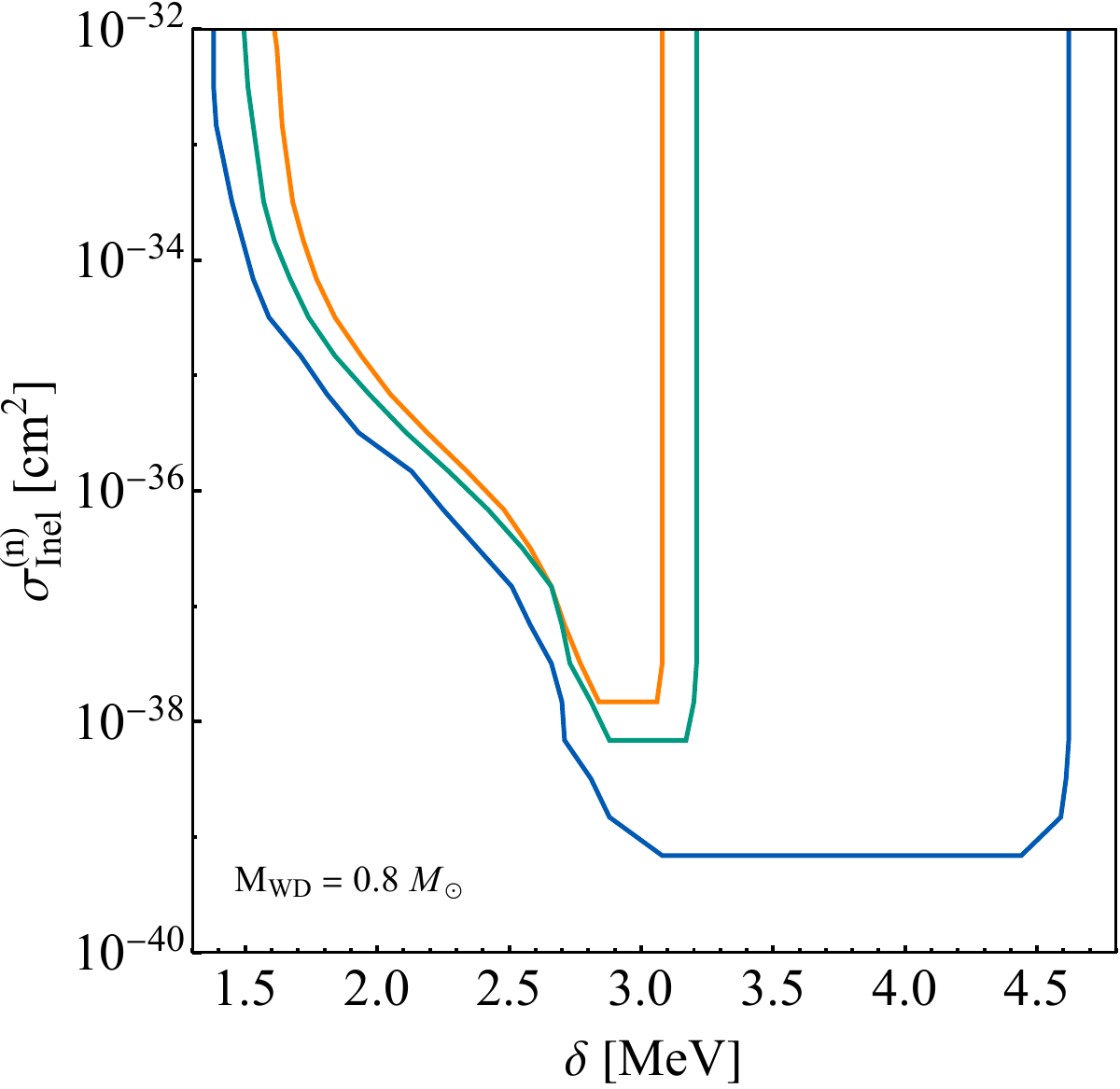}
     \caption[]{Limits on IDM-nucleon inelastic cross-section for $m_\chi = 1 \, \rm TeV$ derived from H.E.S.S. observations of the Galactic Center. The colors indicate the annihilation channel. Each panel corresponds to a different white dwarf benchmark, as specified. For concreteness, we show these results assuming a $\gamma = 1.25$ for the IDM's density profile towards the inner galaxy, a nucleon elastic cross-section $\sigma^{(n)}_{\rm Elas} \simeq 10^{-52} \, \rm cm^2$ and an annihilation cross-section $\langle \sigma v \rangle= 8 \times 10^{-27} \ \rm cm^3/s$.}
    \label{fig:mod_ind_lim}
\end{figure*}

For completeness, we comment on the case where the IDM annihilates to intermediate states that escape the white dwarf's volume, either because they are sufficiently boosted or long-lived. In this scenario, no external annihilation would be required to obtain a finite flux. While this removes the angular momentum threshold for incoming IDM particles ($i.e.$, orbital decay can be ignored), the annihilation rate would remain dependent on the mass splitting through the capture rate itself. As detailed above, the inelastic threshold implies that for increasing mass splitting, a decreasing volume of the white dwarf can capture IDM, $cf.$ Eq.~\eqref{eq:flux_main}. Consequently, the annihilation rate eventually vanishes in the limit $\delta \rightarrow \delta_{\rm max}$. Eq.~\eqref{eq:gamma_flux}, on the other hand, would be modified to account for the decay branching ratio of the intermediate state, its photon spectrum per decay, and its probability of survival over the decay baseline $D_{\rm GC}$. 

\subsection{Comparison with H.E.S.S. Observations}
We now compare the gamma-ray flux sourced by IDM in white dwarfs to existing observations of the Milky Way's center performed by H.E.S.S. in the energy range $E_\gamma \sim (0.1 - 62) \ \rm TeV$ \cite{Malyshev:2015hqa}. For a given set of IDM parameters, we consider an inelastic scattering cross-section value to be excluded if the resulting photon flux is greater than the reported flux in any energy bin. This is a conservative procedure, as modeling astrophysical backgrounds and performing a more elaborate statistical analysis would lead to stronger limits. 

Without committing yet to a specific model, we first consider the resulting gamma-ray fluxes for a few representative annihilation channels into quarks, photons, and electrons. We choose these channels to span the range from hard, monochromatic photon lines to softer continuum spectra from hadronic and leptonic final states. For each case, we set the annihilation branching ratio to unity, and compute the spectra using the PPPC 4 DM ID package \cite{Cirelli:2010xx}. In the context of a full model realization, we will consider IDM annihilation into dark photons and electroweak bosons below in Sec.~\ref{sec:IDM_models}. 

Figure~\ref{fig:mod_ind_lim} shows the prospective limits derived from this comparison. These are drawn as a function of mass splitting, for a fixed $m_\chi = 1 \, \rm TeV$, for three separate white dwarf benchmarks centered about $0.7 \, M_\odot$. We have not attempted here to combine these limits using an extended mass distribution for these objects. This would require a fine-grained resolution in white dwarf mass for our calculations and, moreover, their mass distribution is not well-known in our region of interest. We have opted instead to illustrate the variation of these results centered about the average mass obtained from the N-body simulations reported in Ref.~\cite{Panamarev:2018bwq}. We have also fixed the IDM's nucleon-level elastic and annihilation cross-sections to $\sigma^{(n)}_{\rm Elas} = 10^{-52} \ \rm cm^2$ and $\langle \sigma v \rangle = 8 \times 10^{-27} \, \rm cm^3/s$, as well as taken an IDM profile slope $\gamma = 1.25$. We remark that, for the lower bound $\gamma = 1$ we consider on the IDM's profile, sensitivity can be attained for these channels, if the elastic cross-section is set sufficiently low. We have however selected $\gamma = 1.25$ instead, as a representative value that yields sensitivity across all three channels shown for a range of elastic cross-sections as wide as possible.

The dependence of the inelastic cross-section in Fig.~\ref{fig:mod_ind_lim} with mass splitting reflects the effect of the variable optical depth for both capture and orbital decay: at low mass splittings, our calculations only account for the tiny fraction of IDM particles undergoing a relatively large number of scatters with scraping trajectories at the surface. Consequently, a large cross-section is required to capture sufficient IDM and source a signal comparable to what H.E.S.S. measures. As the mass splitting increases, capture (and partial thermalization) requires less scatters. At the same time, this process also occurs deeper within the object, where the density is higher and a relatively wider range of angular momenta are permitted ($cf.$ Fig.~\ref{fig:theta_min}). Therefore, the cross-section sensitivity improves in this regime. This tendency breaks down, however, when capture occurs deep enough within the white dwarf. The resulting orbits traverse the densest regions of the object, leading to a substantial orbital decay rate. Combined with a progressively smaller effective volume for capture, this eventually suppresses the signal for a sufficiently large $\delta$. 

We comment on the variation of these lines with the IDM mass. Heavier IDM tends to produce more energetic photons from annihilations, which would have less background to compete against. However, the increased mass leads to lower annihilation rates, since the particle flux traversing white dwarfs is also reduced. For our simple method to derive the lines in Fig.~\ref{fig:mod_ind_lim}, we find this reduction to the annihilation rate ultimately suppresses sensitivity for IDM heavier than $\sim 5 \ \rm TeV$. Lower mass IDM, on the other hand, produces less energetic photons, which populate bins with a higher observed flux. Although in this case the rate of IDM passing through white dwarfs is enhanced, the larger background suppresses sensitivity. Moreover, our semi-analytic approach requires by construction $m_\chi \gg m_N$, and these results cannot be safely extrapolated once the IDM mass is $m_\chi \lesssim 100 \ \rm GeV$. Overall, we find sensitivity to this external annihilation effect to be optimal for IDM at or near the TeV scale. 

Finally, we emphasize that there are several existing or proposed astrophysical searches for IDM, such as neutron star and white dwarf heating \cite{McCullough:2010ai,Baryakhtar:2017dbj,Bell:2018pkk,Alvarez:2023fjj}, cosmic-ray boosting in active Galactic nuclei \cite{Gustafson:2025dff}, orbital decay of celestial bodies near Sgr.~A$^\star$ \cite{Acevedo:2025rqu}, and gravitational slingshots from black hole binaries \cite{Acevedo:2026xol,Acevedo:2026tur}. 
However, none currently probe the MeV-splitting regime at this cross-section level and IDM mass scale. Neutron stars are in principle more sensitive targets owing to their higher densities. However, kinetic heating searches in these objects have yet to yield robust constraints. These critically depend on the existence of extremely old, nearby radio-pulsars, or else rely on the uncertain dark matter content in globular clusters. Moreover, dark matter heating effects can be masked by internal heat sources of the neutron star, such as vortex creep motion \cite{Fujiwara:2023tmr}. On the other hand, external annihilation in neutron stars (analogous to the effect explored in this work) is restricted to substantially larger mass splittings than considered here \cite{Acevedo:2024ttq}. 

\section{Model Realizations}
\label{sec:IDM_models}
We have conducted so far most of our analysis in a model-independent manner, with the only assumptions being a suppressed relic abundance of the heavy state $\chi_2$ relative to the light state $\chi_1$, a large tree-level inelastic cross-section, and a much tinier loop-level elastic cross-section. We now illustrate the applicability of this search to concrete models. Specifically, we will consider inelastic dark photon-mediated and (nearly pure) higgsino dark matter. We analyze how these models meet such conditions, and estimate the parameter space that can be probed in each case. 

\subsection{Dark Photon-mediated IDM}
In its most minimal inelastic realization, a small Majorana mass is introduced through some symmetry-breaking scalar field, leading to a Lagrangian density of the form \cite{Batell:2009vb,Zhang:2016dck,Alvarez:2019nwt}
\begin{align}
    \label{eq:dp_lag}
    \mathcal{L} & \supset |D_{\mu}\phi|^2 + V(\phi) - \frac{1}{4} V_{\mu \nu}^2 + \kappa V_{\mu}\partial_{\nu}F^{\mu \nu} \\ & - \frac{1}{2} m_V^2 V_\mu^2  + \bar{\chi}\left(i D_{\mu}\gamma^{\mu}-m_{\chi}\right)\chi + \left(\lambda_D \phi \chi^T C^{-1}\chi + {\rm h.c.} \right) \nonumber
\end{align}
where $V_\mu$ ($V_{\mu\nu}$) is the new $U(1)_D$ gauge boson (field strength tensor) which mixes with the SM photon through a kinetic mixing parameter $\kappa$, $D_{\mu} = \partial_{\mu} + i e_D V_{\mu}$, where $e_D$ is the charge of the dark matter under $U(1)_D$, and $C$ is the charge conjugation operator. We assume the dark photon mediator is lighter than the lightest dark matter state, and so this theory belongs to the secluded model class \cite{Pospelov:2007mp}. In this scenario, a mass splitting $\delta = \lambda_D v_\phi$ is induced when the scalar $\phi$ acquires a vacuum expectation value $v_\phi$. 

We now proceed to verify the above assumptions:

\begin{itemize}
    \item \textit{Relic population:} For MeV-scale splittings, the heavy state may decay into electron-positron pairs, or undergo radiative transitions into the lightest state. The inverse rates of either process in this regime are much shorter than the age of the Universe \cite{Batell:2009vb}. Therefore, we expect the relic abundance to be predominantly populated by the lightest state. This precludes direct detection through exothermic scattering. 
    
    \item \textit{Cross-section:} When the kinematic threshold is met, inelastic scattering into the heavier state proceeds with a reference cross-section \cite{Batell:2009vb}
\begin{align}
   \ \ \ \ \ \ \ \sigma_{\chi p}^{\rm inel} & \ =  \frac{16 \pi \alpha \alpha_D \kappa^2 m_p^2}{m_V^4} \\
   \simeq & \ 1.2 \times 10^{-37}\ {\rm cm^2} \left(\frac{\kappa}{10^{-4}}\right)^2 \left(\frac{\alpha_D}{0.1}\right)\left(\frac{{\rm GeV}}{m_V}\right)^4~, \nonumber
\end{align}
at tree-level, where $\alpha_D = e_D^2/4\pi$. By contrast, the per-nucleon loop-level elastic cross-section is 
\begin{align}
\ \ \ \ \ \ \ \sigma_{\chi p}^{\rm elas} & \ =  \frac{\alpha^2 \alpha_D^2 \kappa^4 m_p^4 f_q^2}{\pi m_V^6}  \\
    \simeq & \ 5.1 \times 10^{-53} \ {\rm cm^2} \left(\frac{\kappa}{10^{-4}}\right)^4 \left(\frac{\alpha_D}{0.1}\right)^2  \left(\frac{{\rm GeV}}{m_V}\right)^6~, \nonumber
\end{align}
where $f_q \sim \mathcal{O}(0.1)$ is a hadronic matrix element \cite{Bramante:2016rdh}. For these example parameters, elastic scattering lies far beyond current direct detection sensitivity, see $e.g.$ Ref.~\cite{XENON:2023cxc,LZ:2024zvo,PandaX:2024qfu}. 
 
\item \textit{Annihilation:} Captured IDM will annihilate into pairs of dark photons, which subsequently decay and produce charged particles. 
In the $m_V \sim 1 - 100 \ \rm GeV$ mass range we consider, the dark photons lack the lifetime or the boost necessary to escape the stellar volume prior to decaying. Therefore, only dark photons produced from annihilations outside the host object will source a potentially observable signal. 
\end{itemize}

In what follows, we fix $\alpha_D$ to the value compatible with relic abundance assuming the standard freeze-out scenario \cite{Pospelov:2007mp},
\begin{eqnarray}
    \alpha_D \simeq 0.037 \, \left(\frac{m_\chi}{\rm 1 \,TeV}\right)~.
\end{eqnarray}
For a fixed IDM mass, all the cross-sections are therefore regulated by the dark photon mass $m_V$ and kinetic mixing parameter $\kappa$. Below, we present our findings in terms of these parameters.

\begin{figure}[t]
    \centering
    \vspace*{0.4cm} 
    \includegraphics[width=\linewidth]{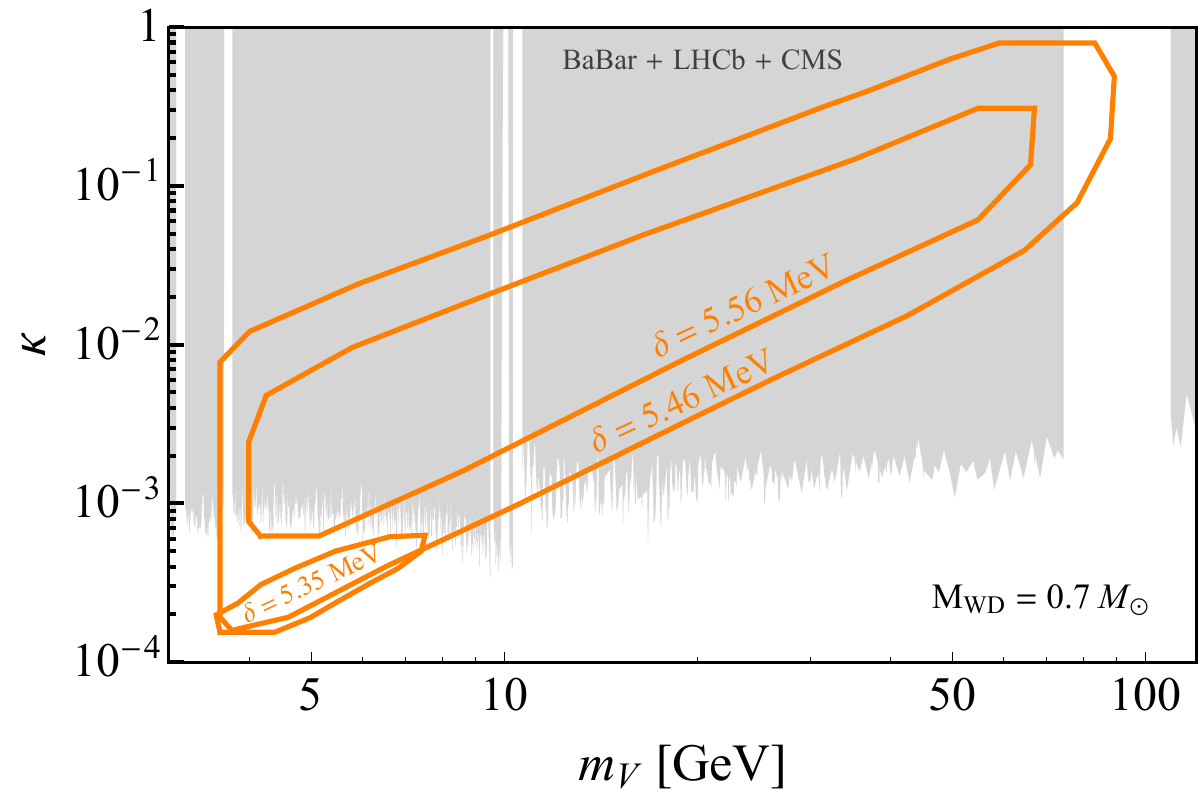}
    \caption{Sensitivity to dark photon-mediated IDM derived from H.E.S.S. observations of the Galactic Center, assuming a monochromatic $0.7 \, M_\odot$ white dwarf mass spectrum and a $\gamma = 1.25$ IDM profile slope. Note that this only accounts for leptonic decays. Each contour corresponds to a different mass splitting, as specified, for $m_\chi = 1 \, \rm TeV$ and dark interaction strength $\alpha_D$ fixed by the required relic abundance. The gray region indicates constraints on the kinetic mixing from various accelerator probes \cite{BaBar:2014zli,LHCb:2019vmc,CMS:2019buh}.}
    \label{fig:dp_lines}
\end{figure} 

The resulting spectral flux is now given by the combination of the various decay channels,
\begin{equation}
    \frac{d\Phi_{\gamma}}{dE_\gamma} = \frac{\mathcal{A}_{\rm WDs}}{4\pi D_{\rm GC}^2} \sum_{f} 2\, {\rm Br}({V \rightarrow f \bar{f}}) \, \frac{dN_{{\gamma}}}{dE_{\gamma}}\bigg|_{f \bar{f}} ~,
\label{eq:dp_gamma_flux}
\end{equation}
where $f$ is any charged fermion within threshold (we do not include decays to $W^{\pm}$, as $m_V \lesssim 100 \ \rm GeV$ throughout this analysis). Relative to Eq.~\eqref{eq:gamma_flux}, an additional factor of two is introduced, accounting for the production of two dark photons per annihilation event, and now Br stands for the decay branching ratio of dark photons into the charged SM state $f$. The produced quarks, in particular, will rapidly hadronize and produce secondary photons from subsequent decays. However, for simplicity, we only consider decays to charged-lepton final states. The inclusion of hadronic decay channels requires a more involved computation beyond the present scope. Our aim here is to illustrate the sensitivity to this model rather than to establish precise limits. Furthermore, neglecting the hadronic contribution to the spectrum is a conservative approximation, since hadronic decays represent a sizable fraction of dark photon decays in this regime, and are expected to increase the photon yield through the production of neutral mesons, particularly from $\pi^0$ decays. 

Figure~\ref{fig:dp_lines} shows the prospective sensitivity to this model for TeV IDM, as contours of dark photon mass and kinetic mixing. We have assumed for simplicity a monochromatic mass spectrum for the white dwarf population centered at $0.7 \, M_\odot$ \cite{Panamarev:2018bwq}, and an IDM profile slope $\gamma = 1.25$. Existing constraints on the kinetic mixing from BaBar, LHCb and CMS are shown in grey for comparison. We find sensitivity only for a narrow splitting range of $\sim 200 \ \rm keV$, close to the corresponding $\delta_{\rm max}$ for the assumed white dwarf benchmark. The lower edge of each contour is set by the minimum inelastic cross-section for which the capture rate is large enough for the resulting photon flux to exceed the observed flux. Above the upper edge, the elastic cross-section becomes large enough that the orbits of captured particles decay too rapidly, suppressing the signal. The contours move to larger $\kappa$ as the dark photon mass increases, since both cross sections fall steeply with $m_V$. In this limit, the sensitivity window eventually closes due to the vanishing inelastic cross-section, which suppresses the capture rate. Conversely, for decreasing $m_V$, the contours move to lower $\kappa$, and the sensitivity window eventually closes as the elastic cross-section grows too large, suppressing the signal through fast orbital decays of the particles. As mentioned above, the sensitivity could be improved if hadronic decays were included in the analysis. 

\subsection{Higgsinos}
An especially well-motivated realization of IDM is nearly pure higgsino dark matter. Higgsinos are the fermionic superpartners of the neutral and charged Higgs bosons in supersymmetric extensions of the SM \cite{Jungman:1995df,Arkani-Hamed:2006wnf}. In the pure-higgsino limit, the lightest neutral states form a nearly degenerate pseudo-Dirac pair, $\tilde{H}_{1,2}$, split by a small mass difference generated through mixing with heavier electroweakinos, see $e.g.$ Ref.~\cite{Fritzsche:2002bi}.

As before, we verify the conditions we have assumed throughout:
\begin{itemize}
    \item \textit{Relic population:} In the nearly degenerate limit \cite{Hall:2011jd,Fox:2014moa}, higgsinos decay into the lightest state either at tree-level producing SM fermions $\tilde{H}_2 \rightarrow \tilde{H}_1 \bar{f}f$ if the mass threshold is met, or at loop-level producing a photon $\tilde{H}_2 \rightarrow \tilde{H}_1 \gamma$. The inverse rates for these decays are much shorter than the age of the universe for the MeV-scale mass splittings we consider, see $e.g.$ Ref.~\cite{Krall:2017xij}. We thus expect this IDM to predominantly be in its lightest state. As with the prior example, exothermic scattering in direct detection experiments will be severely suppressed in this scenario. 
    
    \item \textit{Cross-section:} Inelastic scattering proceeds through an off-diagonal coupling to the $Z$ boson. Neglecting the form factor, this process has a cross-section at the nuclear level \cite{Krall:2017xij}
      \begin{align}
      \label{eq:cs_inel}
   \ \ \ \ \ \ \ \sigma_{\tilde{H} N}  & \ =  \frac{G_f^2 \, \mu_{\tilde{H} N}^2}{8 \pi} \left[(A-Z) - \left(1-4 s_W^2 \right)Z\right]^2 \\
    & \ \ \ \ \ \ \ \simeq (0.8 \ - \ 2.5) \times 10^{-35} \ \rm cm^2~. \nonumber
    \end{align}
    where $G_f \simeq 1.16 \times 10^{-5} \ \rm GeV^{-2}$ is the Fermi coupling, and $s_W^2 \simeq 0.23$ is the sine of the Weinberg angle. The lower line corresponds to scattering against carbon ($A = 12, \, Z=6$) and oxygen ($A = 16, \, Z=8$) in the limit the higgsino is much heavier than the target. Higgsinos may also elastically scatter through electroweak boson loops. However, due to amplitude-level cancellations, the cross-section for this process lies below the neutrino floor of direct detection experiments \cite{Hill:2013hoa,Chen:2019gtm}. For instance, Ref.~\cite{Chen:2019gtm} predicts a higgsino-nucleon cross-section $\sim 3 \times 10^{-50}~{\rm cm^2}$ for the central value of the parton distribution functions used, although it could be signficantly lower than this value (see below). Therefore, as with the dark photon-mediator model above, loop-level elastic scattering lies beyond the current sensitivity of direct detection experiments. 
    
    \item \textit{Annihilation:} Annihilation mainly produces pairs of $W^{\pm}$ and $Z$ bosons. These final states promptly decay, producing photons, neutrinos, and charged particles. Higgsinos may also source monochromatic gamma-ray lines through one-loop processes into $\gamma \gamma$ and $\gamma Z$, which are heavily suppressed compared to the above. While this line feature may provide a cleaner observational signature than the broad continuum emission in a more complete analysis, we do not consider it here for simplicity. Once again, only annihilation that proceeds outside the white dwarf's volume may source a detectable signal.
\end{itemize}

In what follows, we fix the mass $m_{\tilde{H}} \simeq 1.1 \, \rm TeV$, the value at which higgsinos account for the full relic abundance via thermal freeze-out from s-wave annihilation \cite{Arkani-Hamed:2006wnf}. Unlike the dark photon scenario above, the remaining parameters are fixed by the electroweak couplings. Therefore, whether sufficient sensitivity is attained to probe this model depends on the annihilation rate being large enough at the predicted inelastic cross-section. We compute the spectral flux for this model using Eq.~\eqref{eq:gamma_flux}, with the sum running over $W^{\pm}$ and $Z$ bosons as the final states. We set the annihilation branching fractions to ${\rm Br}(\tilde{H} \tilde{H} \to ZZ) \simeq 0.46$ and ${\rm Br}(\tilde{H} \tilde{H} \to W^+ W^-) \simeq 0.54$ \cite{Olive:1990qm}.

\begin{figure}[t]
    \centering
    \vspace*{0.4cm} 
    \includegraphics[width=\linewidth]{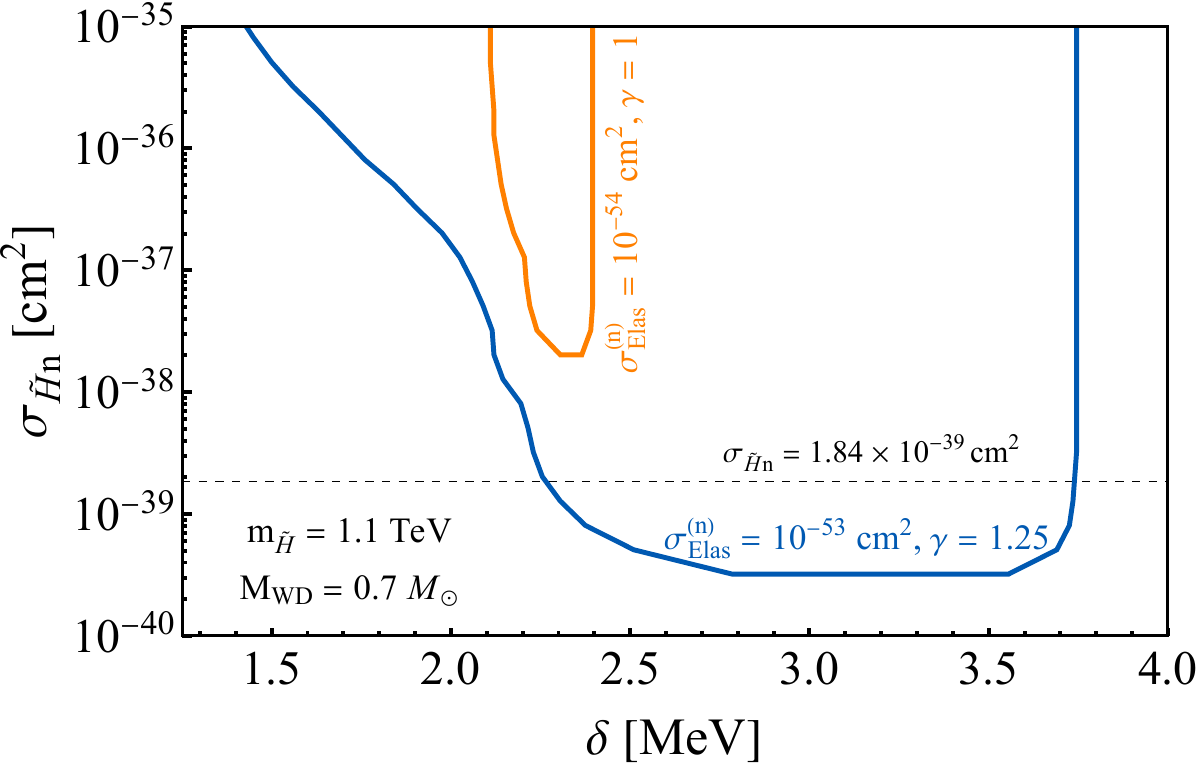}
    \caption{Sensitivity to higgsino-nucleon inelastic cross-section derived from H.E.S.S. observations of the Galactic Center, assuming a monochromatic $0.7 \, M_\odot$ white dwarf mass spectrum. Each curve corresponds to a different IDM profile slope and loop-level elastic cross-section, as specified. The horizontal dashed line indicates the approximate target cross-section.}
    \label{fig:higgs_lines}
\end{figure} 

Figure~\ref{fig:higgs_lines} shows the prospective sensitivity to the higgsino-nucleon inelastic cross-section, as a function of mass splitting. We have assumed again, for simplicity, a monochromatic mass spectrum centered around $0.7 \, M_\odot$ \cite{Panamarev:2018bwq}. The horizontal line indicates the approximate predicted value of the higgsino-nucleon cross-section. Sensitivity is attained whenever the lines cross this threshold.  
We show two benchmark cases corresponding to $\gamma = 1$ and $\gamma = 1.25$, and different loop-level elastic cross-section values. These combinations are meant to illustrate the approximate range of IDM profiles and loop-level cross-sections for which higgsinos would substantially annihilate outside white dwarfs. In general, we require an IDM profile slope $\gtrsim 1$ and elastic cross-sections $\lesssim 10^{-53} \ \rm cm^2$ to attain sensitivity in the $\sim 2 - 4 \ \rm MeV$ splitting range. The required elastic cross-section, in particular, lies below the canonical electroweak-loop prediction, but remains well-motivated in light of recent studies highlighting cancellations that can strongly suppress the elastic scattering rate \cite{Chen:2019gtm,Bisal:2026hpm}. We additionally reiterate that we have been conservative when comparing the flux against H.E.S.S., and a more complete statistical analysis would likely enable sensitivity for larger elastic cross-sections.

\section{Conclusions}
\label{sec:conclusions}
We have analyzed IDM annihilation around white dwarfs in the innermost region of the Galactic Center. Owing to their deep gravitational potential and high density, these objects can focus large amounts of IDM even for relatively weak interaction strengths. Collectively, they provide a boost to the annihilation rate relative to its halo counterpart over a large galactic volume, for a broad range of parameter space.

We have focused on the regime where incoming IDM particles are captured and rapidly thermalized until their kinetic energy falls below the inelastic threshold. This requires a large inelastic scattering cross-section, which guarantees a rapid initial thermalization, and a small elastic scattering cross-section, which extends the remaining thermalization process over long timescales. While these assumptions may appear restrictive, they can be accommodated at present by well-motivated models in which inelastic scattering proceeds at tree-level but elastic scattering is loop-suppressed. Moreover, they permit a semi-analytic treatment of the IDM orbits and their occupation number in terms of the energy and angular momentum at the time inelastic scattering becomes forbidden by kinematics. 

Once inelastic scattering becomes forbidden, a small fraction of the IDM retains sufficient angular momentum to remain on orbits extending beyond the white dwarf, permitting annihilation to occur outside before this population is depleted through the secular energy loss induced by elastic scattering. Because annihilation can proceed externally, we are able to consider arbitrary final states, in contrast with most prior analyses that rely on long-lived or boosted states to transport the energy out of the stellar volume. We have determined the external annihilation rate arising from the equilibrium between capture, annihilation and orbital decay processes. 

Based on H.E.S.S. observations of gamma-ray emission within the central $\sim 18 \, \rm pc$ ($\sim 0.13^\circ$) of the Milky Way, we have derived putative constraints on the inelastic scattering cross-section as a function of mass splitting for IDM masses at the TeV scale. These results are conservative in that we have only required the collective annihilation flux from the white dwarf population to exceed the observed flux at any given energy bin. We find this procedure to be sensitive to a variety of annihilation channels, even under conservative assumptions regarding the dark matter distribution in the central parsec of the Milky Way. We emphasize that the cross-sections and splittings probed are currently inaccessible to direct detection experiments searching for halo dark matter, for two reasons: the kinematic threshold for inelastic scattering is too restrictive for these particles and, for MeV-scale splittings, only the lightest state generically survives as a relic today, precluding threshold-free exothermic scattering as well.

As a concrete illustration, we have applied this framework to well-motivated realizations of IDM: secluded dark photon-mediated dark matter and nearly pure higgsinos. For each scenario, we have identified the approximate parameter space for which the assumptions underlying our treatment are valid, and translated the resulting sensitivity into the corresponding model parameters. Interestingly, we find some of the parameter space of these models can be probed with this search, although we caution about the high uncertainty on the total number of white dwarfs populating this region, which may impact these estimates.

While our analysis has been limited to a regime that enables semi-analytic calculations, more general predictions could be obtained under a fully numerical treatment of the IDM's energy loss and its orbital evolution. Such a study, possibly in combination with more elaborate data comparisons and the consideration of messengers other than gamma-rays, is left for future work. \\  

\section*{Acknowledgements}
I am grateful to Joe Bramante, Rebecca Leane, Michael Peskin, Aidan Reilly, Adam Ritz, Nick Rodd and Linda Xu for helpful discussions. This research was supported in part by the U.S. Department of Energy under Contract DE-AC02-76SF00515. This research was also undertaken thanks in part to funding from the Natural Sciences and Engineering Research Council of Canada through the Arthur B. McDonald Canadian Astroparticle Physics Research Institute. \\ 

\appendix

\section{White Dwarf Structure}
\label{app:WD_struc}
We derive the stellar profile for each benchmark from the Tolman-Volkoff-Oppenheimer hydrostatic equilibrium equation 
\begin{equation}
    \frac{dP}{dr} = - \frac{\rho \, G M}{r^2} \left(1+\frac{P}{\rho}\right)\left(1+\frac{4\pi P r^3}{M}\right) \left(1-\frac{2GM}{r}\right)^{-1}~,
    \label{eq:tov1}
\end{equation}
\begin{equation}
    \frac{dM}{dr} = 4\pi r^2 \rho~.
    \label{eq:tov2}
\end{equation}
Above, $P$ is the pressure, $\rho$ is the energy density, and $M$ is the enclosed stellar mass within a shell of radius $r$. Eqs.~\eqref{eq:tov1} and \eqref{eq:tov2} must be complemented with a suitable equation of state for the white dwarf medium of the form $P = P(\rho)$, and then integrated starting from a central density $\rho(0) = \rho_c$ until $P(r = R_{\rm WD}) = 0$ is reached at an a priori unknown $R_{\rm WD}$. The total white dwarf mass is then $M_{\rm WD} = M(r = R_{\rm WD})$.

We adopt a Feynman-Metropolis-Teller equation of state \cite{2011PhRvC..83d5805R,Rotondo:2011zz}. For the purposes of computing the profiles, we assume pure carbon composition for simplicity. However, we note that the variation of the resulting profiles with the carbon-to-oxygen ratio is negligible \cite{Rotondo:2011zz}. 

\begin{figure}[t]
    \centering
    \hspace*{-0.34cm} \includegraphics[width=\linewidth]{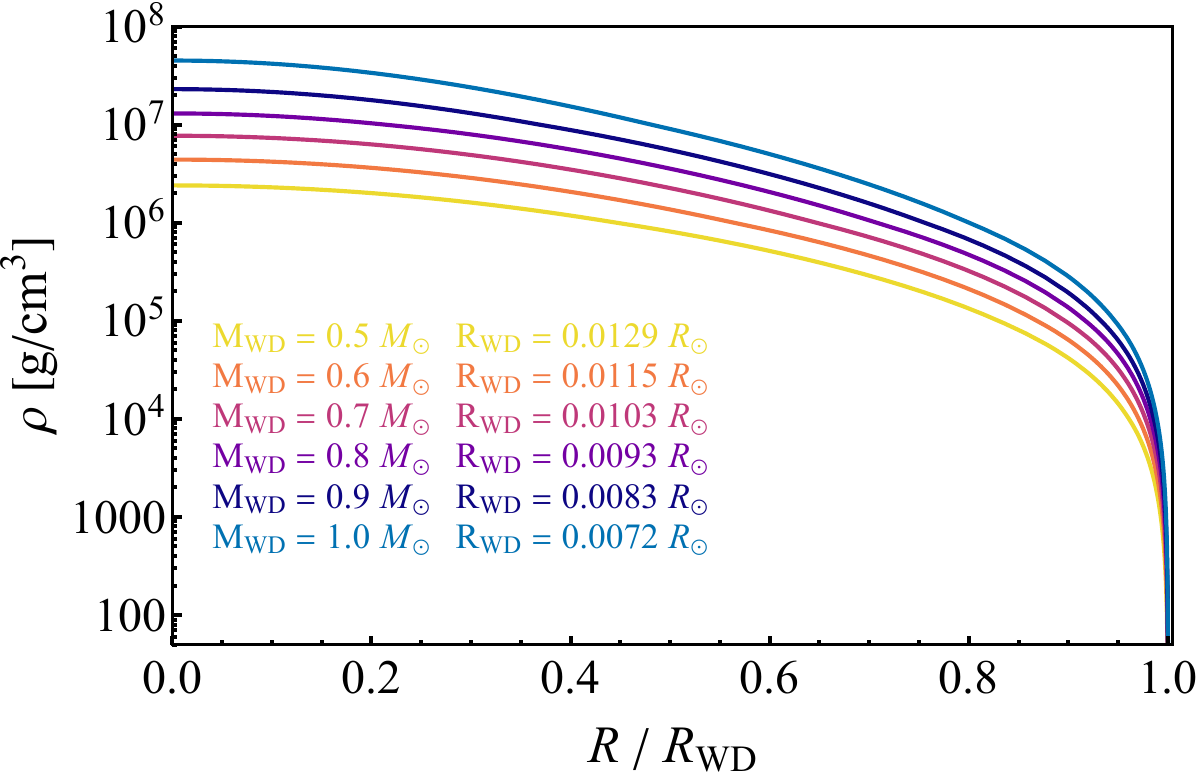}\\[1em] 
    \includegraphics[width=0.97\linewidth]{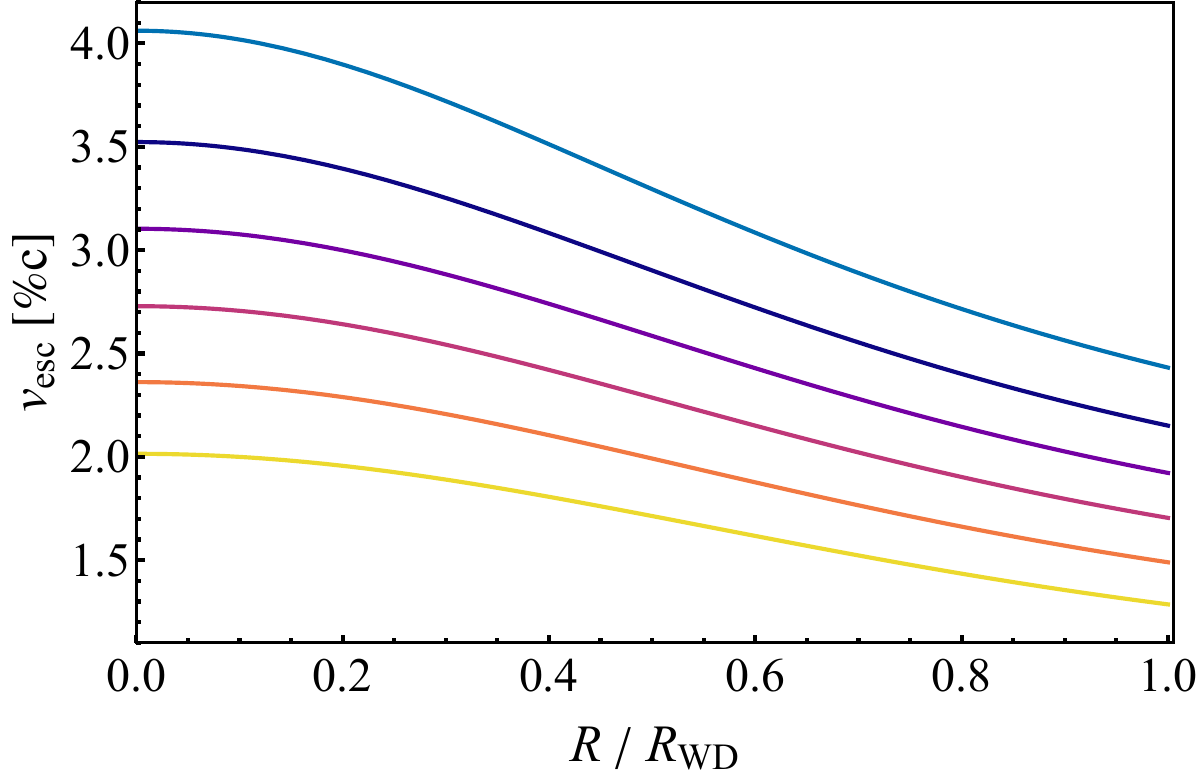}        
    \caption{\textit{Top:} Benchmark white dwarf mass density profiles for masses ranging $0.5 - 1 \ M_\odot$. For reference, their radii are also specified for each case. \textit{Bottom:} Escape velocity profile for the same benchmarks.}
    \label{fig:wd_struc}
\end{figure}

Figure~\ref{fig:wd_struc} shows the resulting density and escape velocity profiles utilized in our calculations, where each color denotes a different benchmark binned by $0.1 \, M_\odot$ increments. The gravitational potential profile can be derived from $\phi(r) = -v^2_{\rm esc}(r)/2$, and is trivially extended as $\phi(r) = - G M_{\rm WD}/r$ outside, though we have not plotted this quantity for brevity.

\newpage

\bibliographystyle{apsrev4-1}
\bibliography{HiggsiWD}
\end{document}